\documentclass[%
 aip,
 jmp,%
 amsmath,amssymb,
 reprint,%
]{revtex4-1}

\usepackage{graphicx}
\usepackage{dcolumn}
\usepackage{bm}
\usepackage{multirow}
\usepackage{color}

\begin{document}

\preprint{AIP/123-QED}

\title[]{Effects of a small magnetic field on homoclinic bifurcations in a low-Prandtl-number fluid
}

\author{Arnab Basak}
\author{Krishna Kumar}%
 \email{kumar.phy.iitkgp@gmail.com}
\affiliation{ 
Department of Physics, Indian Institute of Technology, Kharagpur-721302, India
}%


\date{\today}

\begin{abstract}
Effects of a uniform magnetic field on homoclinic bifurcations in Rayleigh-B\'{e}nard convection in a fluid of Prandtl number $Pr = 0.01$ are investigated using direct numerical simulations (DNS). A uniform magnetic field is applied either in the vertical or in the horizontal direction. For a weak vertical magnetic field, the possibilities of both forward and backward homoclinic bifurcations are observed leading to a spontaneous merging of two limit cycles into one as well as a spontaneous breaking of a limit cycle into two for lower values of the Chandrasekhar's number ($Q\leq 5$). A slightly stronger magnetic field makes the convective flow time independent giving the possibility of stationary patterns at the secondary instability. For horizontal magnetic field, the $x\leftrightharpoons y$ symmetry is destroyed and neither a homoclinic gluing nor a homoclinic breaking is observed. Two low-dimensional models are also constructed: one for a weak vertical magnetic field and another for a weak horizontal magnetic field. The models  qualitatively capture the features observed in DNS and help understanding the unfolding of bifurcations close to the onset of magnetoconvection.
\end{abstract}

\pacs{47.35.Tv, 47.20.Bp, 47.20.Ky}
\keywords{magnetoconvection, fluid patterns, bifurcations}
\maketitle

\begin{quotation}
Dissipative structures spontaneously appear in several continuum mechanical systems when they are subjected to a uniform external forcing. The dynamics of such structures depends upon the nature of the underlying bifurcations. A dynamical system having symmetrically placed saddle and unstable fixed points in its phase space may show the possibility of either homoclinic or heteroclinic bifurcations. Several fluid dynamical systems show homoclinic gluing of two limit cycles into one. This may lead to the phenomenon of pattern bursting, where a dissipative structure disappears for a finite time and then reappears again for a fixed value of the bifurcation parameter. Fluid patterns in Rayleigh-B\'{e}nard convection in low-Prandtl-number fluids with stress-free boundaries show homoclinic as well as heteroclinic dynamics. A uniform applied magnetic field produces Lorentz force which is likely to affect this behavior significantly. The role of a small magnetic field on such behavior is not known. Direct numerical simulations and low-dimensional models for  magnetoconvection show that a small vertical magnetic field allows homoclinic gluing. However, even a very weak horizontal magnetic field destroys homoclinic gluing, as it breaks the rotational symmetry of the flow structure about a vertical axis. The unfolding of bifurcations near the instability onset is quite different for a vertical magnetic field from that for a  horizontal magnetic field.
\end{quotation}

\section{\label{sec:Intro}Introduction}
Extended dissipative systems often form patterns, when a bifurcation parameter is raised above a critical value. Homoclinic bifurcations are observed in several fluid dynamical systems including liquid crystals~\cite{demeter_kramer_1999,peacock_mullin_2001}, Taylor-Couette flow~\cite{abshagen_etal_2001}, Rayleigh-B\'{e}nard convection in low-Prandtl-number fluids with and without rotation~\cite{maity_etal_2013,pal_etal_2013,dan_etal_2014}. They are also known to occur in biological systems~\cite{zebrowski_baranowski_2003}, electrical systems~\cite{glendinning_etal_2001,roy_dana_2006} and optical systems~\cite{herrero_etal_1998}. A homoclinic bifurcation may lead to the possibility of Shilnikov wiggle~\cite{shil'nikov_1969}, three-frequency quasi-periodic orbits~\cite{gerbogi_ott_yorke_1983,lopez_marques_2000}, spontaneous gluing of two limit cycles into a large one or spontaneous  breaking of a limit cycle into two smaller ones~\cite{peacock_mullin_2001,maity_etal_2013,pal_etal_2013,dan_etal_2014} or homoclinic chaos. This may also lead to the phenomenon of pattern bursting~\cite{kft_1996,kpf_2006,bajaj_etal_2002,maity_kumar_2014}, when a fluid pattern disappears for a finite time and reappears again for a fixed value of the bifurcation parameter.

Thermal convection in a thin fluid layer in the presence of a uniform magnetic field, also known as Rayleigh-B\'{e}nard (RB) magnetoconvection, has been studied both experimentally~\cite{nakagawa57,nakagawa59,fauve_etal_1981,fauve_etal_1984,cioni_etal_2000,ao_2001,bm_2002,yanagisawa_etal_2010} and theoretically~\cite{chandra,proctor_weiss_1982,bc82,meneguzzi87,cb89,podvigina_2010,pk_2012,basak_etal_2014,basak_kumar_2015}.   
The study of RB magnetoconvection is useful for geophysical as well as for astrophysical problems~\cite{glatzmaier_etal_1999,cattaneo_etal_2003,rucklidge06}.
Imposition of a vertical magnetic field is known to delay the onset of convection~\cite{nakagawa57,chandra}. It also affects the thermal flux~\cite{cioni_etal_2000,ao_2001,cb89} significantly. The horizontal magnetic field does not affect the threshold of the primary magnetoconvection but forces the pattern of straight rolls  to align along its direction~\cite{fauve_etal_1981,fauve_etal_1984}. In addition, the onset of time dependent flow is delayed~\cite{yanagisawa_etal_2010}. However, the effects of even a weak magnetic field on the unfolding of bifurcations near the primary instability in a very low-Prandtl-number RB system are not known. Instabilities near the onset of RB magnetoconvection are investigated using direct numerical simulations (DNS) in this work. The effects of small uniform magnetic field on homoclinic bifurcations and related fluid patterns in a fluid of Prandtl number $Pr =0.01$ are studied. A weak vertical magnetic field allows homoclinic bifurcations to exist which leads to the possibility of bursting of fluid patterns. However, even a weak horizontal magnetic field destroys the possibility of a homoclinic gluing. Two low-dimensional models, one for the vertical and other for the horizontal magnetic fields, are also constructed to understand the details of bifurcations near onset.

\section{\label{sec:system}Hydromagnetic System}
A thin horizontal layer of an electrically conducting Boussinesq fluid of thickness $d$, kinematic viscosity $\nu$, thermal expansion coefficient $\alpha$, thermal diffusivity $\kappa$, magnetic permeability $\mu$ and magnetic diffusivity $\lambda$ is uniformly heated from below and uniformly cooled from above to maintain an adverse temperature gradient $\beta$ across the fluid layer. The positive direction of the $z$-axis is taken along the vertically upward direction and the $xy$-plane is assumed to be coincident with the bottom layer of the fluid. We have considered two cases of magnetoconvection: (i) a uniform magnetic field is applied along the vertically upward direction, i.e., $\bm{B_0} = (0,0,B_0)$, and (ii) a uniform magnetic field is applied along the $y$-axis, i.e., $\bm{B_0} = (0,B_0,0)$. As the values of the magnetic Prandtl number $Pm = \nu/\lambda$ for all terrestrial fluids are of the order of $10^{-6}$ or less, $Pm$ is set equal to zero.  The basic state is the stationary conduction state~\cite{chandra}, which is described as:
\begin{eqnarray}
T_{s}(z) &=& T_0 -\beta z,\label{eq:T-cond}\\
\rho_{s}(z) &=& \rho_0 (1+\alpha \beta z),\label{eq:rho-cond}\\
P_{s}(z) &=& P_0 + \rho_0 g \left[d-z + \frac{\alpha \beta}{2} (d^2-z^2)\right] - \frac{|B_0|^2}{2\mu},
\label{eq:P-cond}
\end{eqnarray}
where $\rho_0$ is the fluid density and $T_0$ is the fluid temperature at the lower boundary. The symbols $T_s$, $\rho_{s}$ and $P_{s}$ stand for the temperature, fluid density and pressure field, respectively, in the conduction state. The symbol $P_0$  represents a constant pressure at the upper boundary. As soon as the temperature gradient $\beta$ is raised above a critical value, convection sets in. The fluid velocity $\bm{v}\thinspace(x,y,z,t)$ $\equiv$ 
$[v_1(x,y,z,t),v_2 (x,y,z,t),v_3(x,y,z,t)]$ becomes non-zero. The temperature, density, pressure and magnetic fields in the fluid are modified and given by,
\begin{eqnarray}
T_{s}(z) &\rightarrow& T(x, y, z, t)= T_{s}(z)+\theta (x, y, z, t),\label{eq:T-conv}\\
\rho_{s}(z) &\rightarrow& \rho (x, y, z, t)= \rho_{s}(z) + \delta \rho (x, y, z, t),\label{eq:rho-conv}\\
P_{s}(z) &\rightarrow& P(x, y, z, t) = P_{s}(z) + p(x, y, z, t),\label{eq:p-conv}\\
\bm{B}_0 &\rightarrow& \bm{B}(x, y, z, t) = \bm{B}_0 + \bm{b}(x, y, z, t),\label{eq:b-conv}
\end{eqnarray}
where $\theta\thinspace(x,y,z,t)$, $\delta\rho\thinspace(x,y,z,t)$,  $p\thinspace(x,y,z,t)$ and $\bm{b}$\linebreak $\thinspace(x,y,z,t)$ $\equiv$ $[b_1 (x,y,z,t),b_2 (x,y,z,t),b_3(x,y,z,t)]$ denote the changes in the temperature, density, pressure and the uniform magnetic field due to convection, respectively. Please note that the convective pressure $p$ also includes the change due to induced magnetic field. For Boussinesq fluids, $\delta \rho = \rho_0 \alpha \theta$.  We now put the hydromagnetic equations in dimensionless form by measuring all length scales in units of $d$, time scales in units of the viscous diffusion time $\tau_{vis} = d^2/\nu$, the convective temperature in units of $\beta d \nu/\kappa$, and the induced magnetic field in units of $B_0 \nu/\lambda$.  

\subsection{\label{sec:vertical}Relevant equations with a vertical magnetic field}
The application of a uniform vertical magnetic field is considered first. The magnetoconvection, in Boussinesq approximation, is governed by the following  dimensionless equations:
 \begin{eqnarray}
&\partial_t\bm{v}+(\bm{v}\cdot\nabla)\bm{v}=-\nabla p+\nabla^2\bm{v}+Q\partial_z\bm{b}+Ra\theta\bm{e}_3,\label{eq:mom-v}\\
&\nabla^2\bm{b}=-\partial_z\bm{v},\label{eq:mag-v}\\
&Pr[\partial_t\theta+(\bm{v}\cdot\nabla)\theta]=\nabla^2\theta+v_3,\label{eq:theta}\\
&\nabla\cdot\bm{v}=\nabla\cdot\bm{b}=0,\label{eq:cont}
\end{eqnarray}
where $\bm{e}_3$ is a unit vector directed opposite to the acceleration due to gravity $\bm{g}$. The Rayleigh number $Ra = \alpha \beta gd^4/(\nu \kappa$) is a measure of the buoyancy force, and the Prandtl number $Pr = \nu/\kappa$ is the ratio of the kinematic viscosity to the thermal diffusivity. The Chandrasekhar's number $Q$ is defined as $Q = B_0^2d^2/(\rho_0\nu\mu\lambda)$ $=$ $\sigma B_0^2d^2/(\rho_0\nu)$, where $\sigma = 1/(\mu\lambda)$ is the electrical conductivity of the fluid. Here $Q$, which is a measure of the external magnetic field strength, is a new bifurcation parameter for the problem of magnetoconvection. All dimensional parameters are taken in SI units. The horizontal boundaries located at $z=0$ and $z=1$ are considered to be \emph{stress-free} and thermally conducting. As the induced magnetic field is slaved to the velocity field, the boundary conditions on the induced magnetic fields have to be consistent with those of the velocity fields. The equation for the induced magnetic field (Eq.~\ref{eq:mag-v}) suggests that the expansions of the induced  horizontal magnetic fields $b_1$, $b_2$ should be similar to that of the vertical velocity $v_3$, and the expansion of the induced vertical magnetic field $b_3$ should be similar to those of the horizontal velocities ($v_1$ and $v_2$). A consistent expansion for the induced magnetic field is possible by considering the boundaries to be electrically insulating~\cite{basak_etal_2014}. This leads to the following conditions at the horizontal boundaries located at $z=0$ and $z=1$:  
\begin{equation}
\frac{\partial v_1}{\partial z}=\frac{\partial v_2}{\partial z}=v_3=\theta=b_1=b_2= \frac{\partial b_3}{\partial z}=0. 
\end{equation}

The hydromagnetic system (Eqs.~\ref{eq:mom-v}-\ref{eq:cont}) possesses a four-fold rotational symmetry about a vertical axis. This leads to the invariance of the hydromagnetic system under the following transformations:\\
(1) $v_1 \rightarrow v_2$, $v_2 \rightarrow -v_1$, $b_1 \rightarrow b_2$, $b_2 \rightarrow -b_1$ under an anti-clockwise rotation of the coordinate axes by an angle of $\pi/2$ about the $z$-axis ($x \rightarrow y$ and $y \rightarrow -x$), while other fields remain unchanged.\\ 
(2) Similarly, $v_1 \rightarrow -v_2$, $v_2 \rightarrow v_1$, $b_1 \rightarrow -b_2$, 
$b_2 \rightarrow b_1$ under an anti-clockwise rotation of the coordinate axes by an angle of $3\pi/2$ about the $z$-axis ($x \rightarrow -y$ and $y \rightarrow x$), while other fields do not transform. \\  
(3) $v_1 \rightarrow -v_1$, $v_2 \rightarrow -v_2$, $b_1 \rightarrow -b_1$, 
$b_2 \rightarrow -b_2$ under inversion ($x \rightarrow -x$ and $y \rightarrow -y$). Other fields do not change sign under the inversion symmetry, which is equivalent to a rotation by an angle of $\pi$ about the $z$-axis.

The system is also invariant under reflections about the $x$- and $y$-axes in the $xy$-plane. The reflection symmetry about the $x$-axis ($x \rightarrow x$ and $y \rightarrow -y$) leads to the transformations:  $v_2 \rightarrow -v_2$ and $b_2 \rightarrow -b_2$. The reflection symmetry about the $y$-axis ($x \rightarrow -x$ and $y \rightarrow y$) leads to the transformations:  $v_1 \rightarrow -v_1$, and $b_1 \rightarrow -b_1$.  The system is also invariant under a mirror reflection about the mid-plane ($z = 1/2$ plane). This introduces the following properties: $v_3 \rightarrow -v_3$, $b_1 \rightarrow -b_1$, $b_2 \rightarrow -b_2$ and $\theta \rightarrow -\theta$ as $1/2+z \rightarrow 1/2-z$.

\subsection{\label{sec:horizontal}Relevant equations with a horizontal magnetic field}
For the case of a uniform horizontal magnetic field along the $y$-axis, the equations for momentum transfer and the induced magnetic field [Eqs.~\ref{eq:mom-v}-\ref{eq:mag-v}] are replaced by,
\begin{eqnarray}
&\partial_t\bm{v}+(\bm{v}\cdot\nabla)\bm{v}=-\nabla p+\nabla^2\bm{v}+Q\partial_y\bm{b}+Ra\theta\bm{e}_3,\label{eq:mom-h}\\
&\nabla^2\bm{b}=-\partial_y\bm{v}.\label{eq:mag-h}
\end{eqnarray} 
The other equations~[Eqs.~(\ref{eq:theta}) and (\ref{eq:cont})] remain unchanged. The horizontal bounding surfaces are stress-free and thermally conducting as mentioned earlier. An inspection of Eq.~\ref{eq:mag-h} suggests that the expansions of the induced magnetic fields $b_1$, $b_2$ and $b_3$ on the $z$-coordinate should be like those of $v_1$, $v_2$ and $v_3$, respectively. These requirements are compatible with electrically conducting boundaries in the case of a horizontal magnetic field.
We have therefore considered the horizontal boundaries to be electrically conducting in this case. This leads to the following boundary conditions 
\begin{equation}
\frac{\partial v_1}{\partial z}=\frac{\partial v_2}{\partial z}=v_3=\theta= \frac{\partial b_1}{\partial z}=\frac{\partial b_2}{\partial z}=b_3=0.
\end{equation}
at $z=0$ and $z=1$.

The presence of a horizontal magnetic field alters the symmetries of the  hydromagnetic system (Eqs.~\ref{eq:theta}, \ref{eq:cont}, \ref{eq:mom-h} and \ref{eq:mag-h}). The system loses the four-fold rotational symmetry. The hydromagnetic system is, however, invariant under the inversion symmetry.  This leads to the following transformation properties: $v_1 \rightarrow -v_1$, $v_2 \rightarrow -v_2$ and $b_3 \rightarrow -b_3$ as $x \rightarrow -x$ and $y \rightarrow -y$. Other fields do not change sign under the inversion symmetry. As the horizontal magnetic field is considered along the $y$-axis, the transformation properties of the convective fields under reflection about the $y$-axis are different from those under the reflection about the $x$-axis. Fields $v_1 \rightarrow -v_1$ and $b_1 \rightarrow -b_1$ under a reflection about the $y$-axis, while other fields do not change sign. However, $v_2 \rightarrow -v_2$, $b_1 \rightarrow -b_1$ and $b_3 \rightarrow -b_3$ under reflection about the $x$-axis, while other fields remain unchanged.  The horizontal external field modifies the symmetries of the problem. It is therefore expected to affect unfolding of bifurcations. The transformation properties of the convective fields under reflection about the mid-plane, however, remain the same as in the case of a vertical magnetic field. The boundary conditions considered for the case of a horizontal magnetic field are compatible with these symmetries, and are different from those compatible with the symmetries with a vertical magnetic field.    

\begin{table*}[ht]
  \begin{center}
\def~{\hphantom{0}}
  \begin{tabular}{c|c|c|c|c|c}
\hline
Fluid  &\multicolumn{3}{c|}{Vertical magnetic field}&\multicolumn{2}{c}{Horizontal magnetic field} \\
\cline {2-4} \cline{5-6}
patterns  & $r (Q = 4)$ & $r (Q = 5)$ & $r (Q = 6)$ & $r (Q = 5)$ & $r (Q = 10)$ \\ 
\hline\hline
SR & $\leq 1.003$ & $\leq 1.010$ & $\leq 1.310$ & $-$ & $\leq 1.007$ \\
\hline
PB-I & $1.004 - 1.055$ & $-$ & $-$ & $-$ & $-$ \\
\hline
PB-II & $-$ & $1.011 - 1.076$ & $-$ & $-$ & $-$ \\
\hline
OCR-I & $-$ & $1.077 - 1.169$ & $-$ & $-$ & $-$ \\
\hline
OCR-II & $1.056 - 1.146$ & $1.170 - 1.199$ & $-$ & $-$ & $-$ \\
\hline
OCR-I	& $1.147 - 1.188$ & $1.200 - 1.217$ & $-$ & $-$ & $-$ \\
\hline
OCR-II & $1.189 - 1.204$ & $1.218 - 1.227$ & $-$ & $-$ & $-$ \\
\hline
OCR	& $-$ & $-$ & $-$ & $\leq 1.200$ & $1.008 - 1.300$ \\
\hline
CR	& $1.205 - 1.226$ & $1.228 - 1.240$ & $-$ & $\geq 1.201$ & $\geq 1.301$ \\
\hline
SQ	& $\geq 1.227$ & $\geq 1.241$ & $\geq 1.311$ & $-$ & $-$  \\
\hline
 \end{tabular}
\caption{Fluid patterns near the onset of magnetoconvection observed in direct numerical simulations (DNS) for a fluid of Prandtl number $Pr=0.01$. Stationary straight rolls (SR), periodic bursting (PB-I with $|W_{101}|_{max} = |W_{011}|_{max}$  and PB-II with $|W_{101}|_{max} \neq |W_{011}|_{max}$), a periodic competition between two sets of mutually perpendicular cross-rolls of equal amplitudes (OCR-I), a periodic competition between mutually perpendicular sets of cross-rolls (OCR-II) of unequal amplitudes, simple oscillating cross-rolls (OCR) for horizontal magnetic field,  stationary cross-rolls (CR, $W_{101} \neq W_{011}$) and stationary squares (SQ, $W_{101} = W_{011}$).}
\label{tab:fluid_patterns}
 \end{center}
 \end{table*}

\subsection{\label{sec:DNS}Direct Numerical Simulations}
We perform direct numerical simulations on both the hydromagnetic systems for a low-Prandtl-number fluid ($Pr=0.01$) using pseudo-spectral method.  All the fields are assumed to be periodic in the horizontal plane. Any convective field $f$ then follows an additional symmetry $f(x + 2l\pi/k_c, y + 2m \pi/k_c, z, t)$ $=$ $f(x, y, z, t)$.  The components of the velocity field $\bm{v}\thinspace(x,y,z,t)$, the convective temperature field ${\theta}\thinspace(x,y,z,t)$, and the convective pressure field $p\thinspace(x,y,z,t)$ are expanded as: 
\begin{eqnarray}
{v_1} (x,y,z,t) &=& \sum_{l,m,n} U_{lmn}(t) e^{ik(lx+my)} \cos{(n\pi z)},\label{eq.U}\\
{v_2} (x,y,z,t) &=& \sum_{l,m,n} V_{lmn}(t) e^{ik(lx+my)}\cos{(n\pi z)},\label{eq.V}\\
{v_3} (x,y,z,t) &=& \sum_{l,m,n} W_{lmn}(t) e^{ik(lx+my)} \sin{(n\pi z)},\label{eq.W}\\
{\theta} (x,y,z,t) &=& \sum_{l,m,n} {\Theta}_{lmn}(t) e^{ik(lx+my)}\sin{(n\pi z)}, \label{eq.T}\\
{p} (x,y,z,t) &=& \sum_{l,m,n} {\Pi}_{lmn}(t) e^{ik(lx+my)} \cos{(n\pi z)},\label{eq.P}
\end{eqnarray} 
where $l, m, n$ are integers. All possible choices of these integers satisfying the following equation are possible. 
\begin{equation}
ilk U_{lmn} + imk V_{lmn} + n\pi W_{lmn}=0.
\end{equation}

The relevant symmetry for homoclinic bifurcations is the invariance of the system under the interchange of indices $l$ and $m$ for the modes $W_{lmn}$ and $\Theta_{lmn}$. If $W_{lmn}$ and $\Theta_{lmn}$ describe a fluid pattern for a given set of integers $l$ and $m$, then  $W_{mln}$ and $\Theta_{mln}$ also represent a solution.  In the case of a vertical magnetic field, the expansions for the components of the induced magnetic field $\bm{b}$ are given by,
\begin{eqnarray}
{b_1} (x,y,z,t) &=& \sum_{l,m,n} {\Psi}_{lmn}(t) e^{ik(lx+my)} \sin{(n\pi z)}, \label{eq.bv1}\\
{b_2} (x,y,z,t) &=& \sum_{l,m,n} {\Phi}_{lmn}(t) e^{ik(lx+my)} \sin{(n\pi z)},\label{eq.bv2}\\
{b_3} (x,y,z,t) &=& \sum_{l,m,n} {\Gamma}_{lmn}(t) e^{ik(lx+my)} \cos{(n\pi z)}. \label{eq.bv3}
\end{eqnarray}
As the induced magnetic filed $\bm{b}$ is slaved to the velocity field $\bm{v}$, the Fourier amplitudes ${\Psi}_{lmn}(t)$, ${\Phi}_{lmn}(t)$ and ${\Gamma}_{lmn}(t)$ may be expressed in terms of the complex amplitudes ${U}_{lmn}(t)$, ${V}_{lmn}(t)$ and ${W}_{lmn}(t)$, respectively, using Eq.~\ref{eq:mag-v}. 

As the boundary conditions for the velocity, convective temperature and pressure fields remain unchanged  in the case of a horizontal magnetic field, the expansions for these fields remain the same. The expansions for the components of the induced magnetic field in this case are given by, 
\begin{eqnarray}
{b_1} (x,y,z,t) &=& \sum_{l,m,n} {\Psi}_{lmn}(t) e^{ik(lx+my)} \cos{(n\pi z)},\label{eq.bh1}\\
{b_2} (x,y,z,t) &=& \sum_{l,m,n} {\Phi}_{lmn}(t) e^{ik(lx+my)} \cos{(n\pi z)},\label{eq.bhv2}\\
{b_3} (x,y,z,t) &=& \sum_{l,m,n} {\Gamma}_{lmn}(t) e^{ik(lx+my)} \sin{(n\pi z)}.\label{eq.bh3}
\end{eqnarray}
The Fourier amplitudes ${\Psi}_{lmn}(t)$, ${\Phi}_{lmn}(t)$ and ${\Gamma}_{lmn}(t)$ may be expressed in terms of the complex amplitudes ${U}_{lmn}(t)$, ${V}_{lmn}(t)$ and ${W}_{lmn}(t)$, respectively, using Eq.~\ref{eq:mag-h} in this case. The symmetry of a field under interchange of indices $l$ and $m$ is broken by a horizontal magnetic field. The integration of the two hydromagnetic systems [(\ref{eq:mom-v}-\ref{eq:cont}) and (\ref{eq:theta}-\ref{eq:cont},\ref{eq:mom-h}-\ref{eq:mag-h})] are done on $64\times 64\times 64$ spatial grid points using a standard fourth order Runge-Kutta (RK4) method with a time step $\delta t \le 0.001$. 

\begin{figure*}[ht]
\begin{center}
\resizebox{0.8\textwidth}{!}{%
  \includegraphics{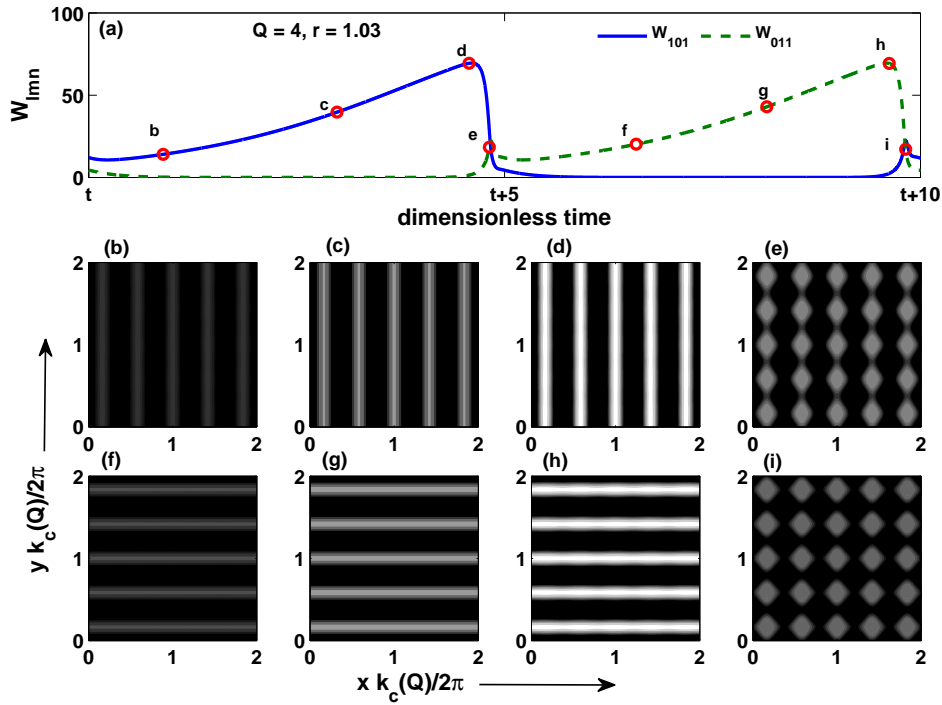}
}
\caption{\label{fig:vert_PB} (a) Time evolution of the two largest Fourier modes $W_{101}$ [continuous blue (black) curve] and $W_{011}$ [broken green (gray) curve], computed from DNS, are shown in the presence of a weak vertical magnetic field [$Q=4$, $k_c(Q)=2.395$] for $Pr=0.01$ and $r=1.03$. The contour plots [(b)-(i)] of the convective temperature field at the mid-plane ($z=1/2$) show the fluid patterns at the instants marked by letters `b' to `i' in part (a), respectively.  Every time the amplitude of a 
growing set of straight rolls becomes large enough, a new set of rolls 
is excited in the direction perpendicular to the old set of rolls.}
\end{center}
\end{figure*}

\section{\label{sec:fluid_patterns}Bifurcations and fluid patterns}
The effects of a small magnetic field (either in the vertical or in the horizontal direction) on the fluid instabilities near the onset of magnetoconvection are investigated in a fluid of $Pr = 0.01$. Table~\ref{tab:fluid_patterns} enlists the various fluid patterns computed from direct numerical simulations (DNS) of Rayleigh-B\'{e}nard magnetoconvection in a fluid of $Pr = 0.01$.

\begin{figure}[h]
\begin{center}
\resizebox{0.45\textwidth}{!}{%
  \includegraphics{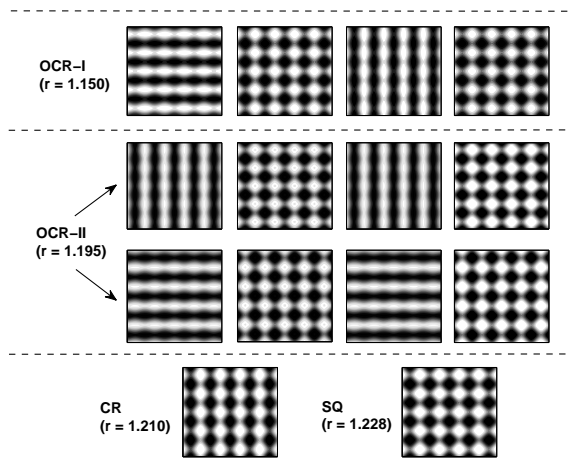}
}
\caption{\label{fig:vert_patterns} Typical fluid patterns obtained from DNS in the presence of a uniform vertical magnetic field for $Q=4$ and $Pr=0.01$.}
\end{center}
\end{figure}

\subsection{\label{sec:patterns_vertical}Fluid patterns with a vertical magnetic field}
We now discuss the fluid patterns observed in the presence of a weak vertical magnetic field obtained from DNS. A pattern of stationary straight rolls appears at the onset of magnetoconvection. The critical value of $Ra_c (Q)$ for the primary instability depends on $Q$. 
The secondary instability occurs soon as the reduced Rayleigh number $r = Ra/Ra_c(Q)$ is raised above $r = r_{O1} > 1$, which is very close to unity. The stationary rolls become unstable, and a time periodic competition between two sets of mutually perpendicular rolls begins at the secondary instability. The Fourier mode $W_{101}$ ($W_{011}$) is the largest mode for a set of rolls parallel to the $y$-axis ($x$-axis). Depending on the value of $Q$, two types of limit cycles (oscillatory solutions) are observed in any quadrant of the $W_{101}-W_{011}$ plane at the secondary instability:\\  
(i) A single limit cycle, a part of which coincides alternately with the $W_{101}$- and $W_{011}$-axes in the $W_{101}-W_{011}$ plane for a finite time. The modes $W_{101}$ and $W_{011}$ are identical with a constant phase difference between them.\\
(ii) Two distinct smaller limit cycles, a part of the first one coincides with the $W_{101}$-axis and a part of the second one coincides with the $W_{011}$-axis for a finite time. Depending on initial conditions, one of the two limit cycles is selected in a simulation. 
 
\begin{figure}[h]
\begin{center}
\resizebox{0.5\textwidth}{!}{%
  \includegraphics{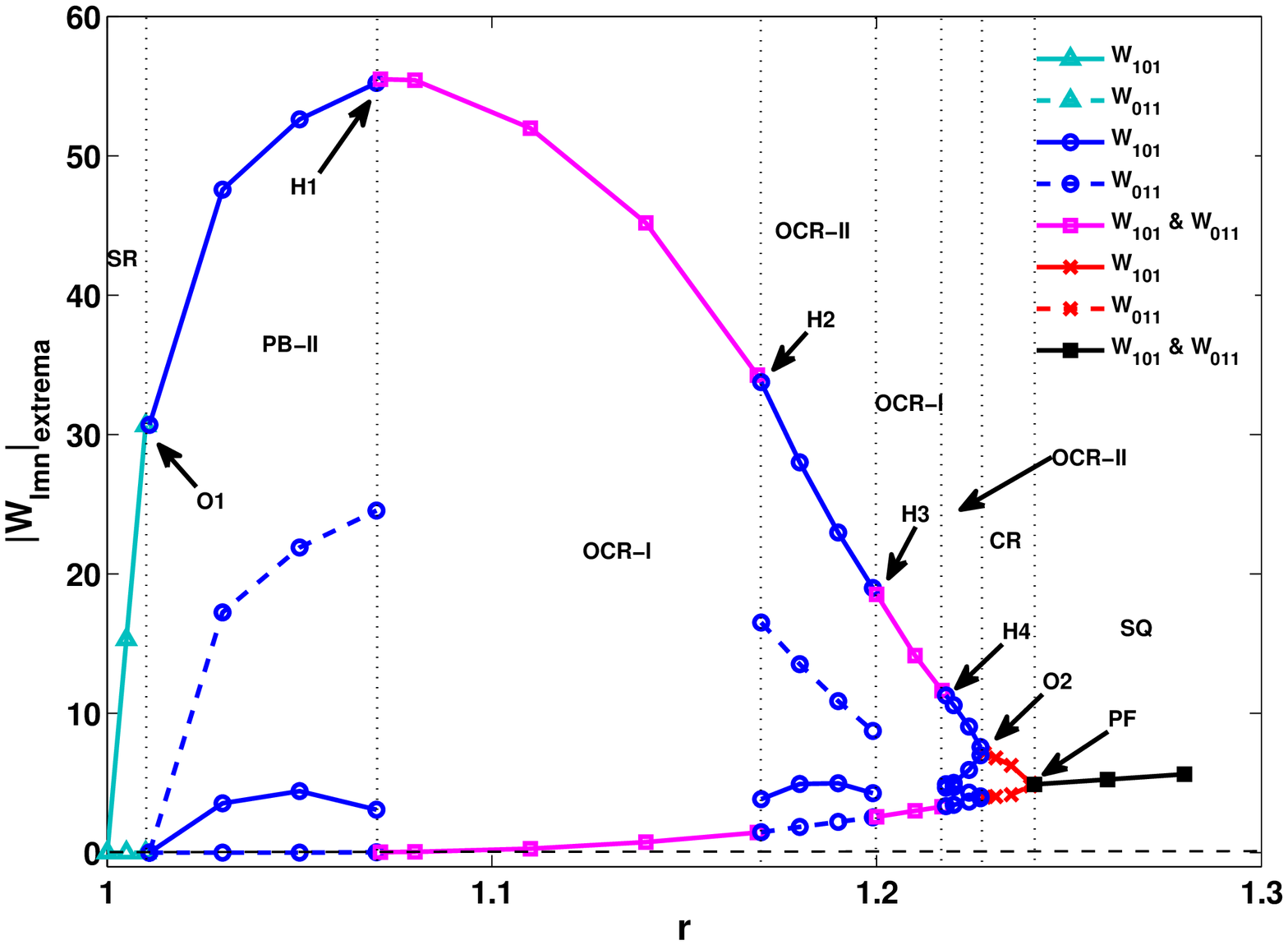}
}
\caption{\label{fig:vert_bif_DNS_Q5} Bifurcation diagram obtained from DNS for $Pr=0.01$ and $Q=5$: The extrema of the Fourier modes $W_{101}$ and $W_{011}$ are plotted as a function of the reduced Rayleigh number $r$ in the case of a vertical magnetic field. Arrows mark the locations of different bifurcation points. The symbols $O1$, $H{j}$ ($j=1,2,3,4$), $O2$, and $PF$ stand for forward Hopf bifurcation at $r=r_{O1}$, four homoclinic bifurcation points at $r=r_{Hj}$, inverse Hopf bifurcation at $r=r_{O2}$ and an inverse pitch-fork bifurcation at $r=r_{PF}$, respectively.
Solid and broken curves show the extrema of the modes $W_{101}$ and $W_{011}$, respectively. The pink curves show the identical values for the extrema of both the modes.}
\end{center}
\end{figure}

Figure~\ref{fig:vert_PB}(a) shows the temporal evolution of the two largest Fourier modes $W_{011}$ and $W_{101}$ for $Q=4$. Both the Fourier modes $W_{101}$ and $W_{011}$ show a relaxation oscillation involving two time scales. The mode $W_{101}$ first grows slowly for a time $\tau_{growth}$ and then rapidly decays to zero in a very short period $\tau_{decay}$ ($<< \tau_{growth}$). Shortly before the mode $W_{101}$ reaches its maximum, the mode $W_{011}$ is excited. Once the mode $W_{101}$ becomes equal to zero, it remains zero for a finite time.  Actually, all two-dimensional modes $W_{l0n}$ ($W_{0mn}$) become zero for a finite time once a set of straight rolls parallel to the $y$-axis ($x$-axis) disappears. The mode $W_{101}$ starts growing again a little before the mode $W_{011}$ reaches its maximum. This is an example of the phenomenon of periodic bursting of patterns in magnetoconvection. The solution corresponding to a single large limit cycle in any quadrant of the $W_{101}-W_{011}$ plane is labeled as PB-I here.

\begin{figure}[h]
\begin{center}
\resizebox{0.5\textwidth}{!}{%
  \includegraphics{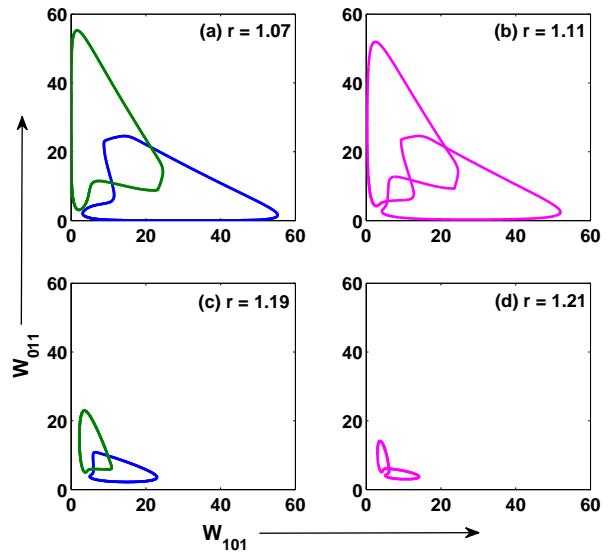}
}
\caption{\label{fig:vert_phase_DNS} Phase portrait computed from DNS for a vertical magnetic field ($Q=5$): (a) Two possible limit cycles [blue (black) and green (gray) colored orbits] for $r =1.07$ showing the phenomenon of bursting (PB-II), (b) a glued limit cycle [orbit in pink (light gray) color] for $r=1.11$ showing a competition between two sets of cross-rolls (OCR-I), (c) two limit cycles (OCR-II) for $r=1.19$, and (d) a glued limit cycle (OCR-I) again at $r=1.21$.}
\end{center}
\end{figure}

Fig.~\ref{fig:vert_PB}(b)-(i) show the contour plots of the convective temperature field at $z=1/2$ at different instants marked by letters `b' to `i' in Fig.~\ref{fig:vert_PB}(a). As the amplitude of a set of straight rolls grows closer to its maximum value, a new set of straight rolls in a direction perpendicular to the previous set of rolls is excited. Shortly after the excitation of another set, the old set of rolls disappears and does not grow for a finite time. Shortly before the amplitude of the new set of rolls reaches its maximum, the old set of rolls is excited once again. This leads to a periodic bursting of mutually perpendicular set of rolls. As $r$ is raised further, the minima of both the modes $W_{101}$ and $W_{011}$, which were zero at the secondary instability, become non-zero. This leads to a periodic competition between two sets of cross-rolls. There are two possibilities:\\
(i)  The two largest modes  $W_{101}$ and $W_{011}$ are identical but with a phase difference between them.  The corresponding limit cycle in the $W_{101}-W_{011}$ plane shrinks in size and does not coincide with any of the axes. The resulting pattern of oscillating cross-rolls is labeled here as OCR-I. A typical pattern is shown in the first row of Fig.~\ref{fig:vert_patterns} for $r=1.150$ and $Q=4$.\\
(ii) The extrema of the two largest modes $W_{101}$ and $W_{011}$ are unequal. 
Two sets of oscillating cross-rolls are possible: one with $|W_{101}|_{max}$ $>$ 
$|W_{011}|_{max}$ and another with $|W_{011}|_{max}$ $>$ $|W_{101}|_{max}$. The corresponding patterns of oscillatory cross-rolls are labeled here as OCR-II. This happens when a single limit cycle  spontaneously breaks into any one of the two possible smaller limit cycles.  Two possible patterns are shown in the second row of Fig.~\ref{fig:vert_patterns} for $r=1.195$ and $Q=4$.\\
As $r$ is raised to higher values keeping the value of $Q$ fixed, each of the two possible limit cycles in each quadrant of the $W_{101}-W_{011}$ plane shrinks to a (fixed) point. This leads to a pattern of stationary cross-rolls (CR) for  $1.205 \leq r \leq 1.226$ and a pattern of stationary squares (SQ) for $ r > 1.226$ and $Q = 4$. The third row of Fig.~\ref{fig:vert_patterns} show patterns of stationary cross-rolls and squares for $r=1.210$ and $r=1.228$, respectively, at a fixed values of $Q=4$. 

\begin{figure}[h]
\begin{center}
\resizebox{0.5\textwidth}{!}{%
  \includegraphics{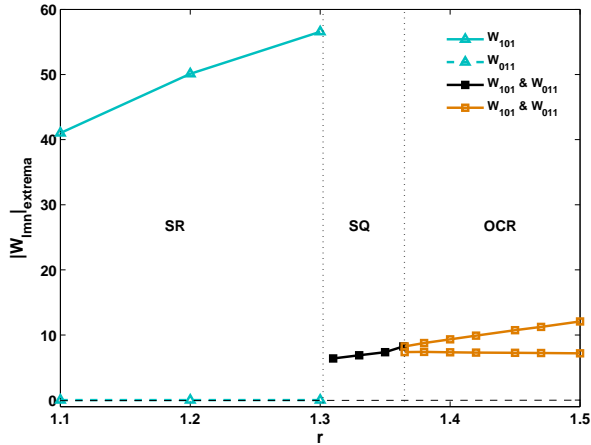}
}
\caption{\label{fig:vert_bif_DNS_Q6} Bifurcation diagram for vertical magnetic field: Plot of $|W_{101}|_{extrema}$ as a function of $r$ obtained from DNS for $Pr=0.01$ and $Q=6$.}
\end{center}
\end{figure}

Figure~\ref{fig:vert_bif_DNS_Q5} shows a bifurcation diagram obtained from DNS for $Q=5$ in the case of a weak vertical magnetic field. Fourier modes $W_{101}$ and $W_{011}$ are plotted as a function of $r$. As soon as $r$ is raised above unity (i.e., $Ra > Ra_c$), the mode $W_{101}$ starts growing but the mode $W_{011}$ does not grow. The solid and broken cyan (light gray) curves show the variation of fixed values of the modes $W_{101}$ and $W_{011}$, respectively, as a function of $r$ for $ 1 < r < 1.011$. In fact all the modes $W_{0mn}$ are zero in this window of $r$. The onset of magnetoconvection is stationary and a set of straight rolls appears at the primary instability.  The symmetry under rotation by an angle $\pi/2$, allows a set of rolls parallel to the $x$-axis ($W_{101}=0$ and $W_{011} \neq 0$). The symmetry of the plane also allows  two similar sets of stationary rolls with the flow direction reversed. There are four sets of roll fixed points in the $W_{101}-W_{011}$ plane. Depending on the initial conditions, one set of stationary rolls is selected in a given simulation. We have observed all four fixed points in the simulations by taking appropriate initial conditions. 

The straight rolls become unstable as $r$ is raised above $r=r_{O1}=1.011$. A forward Hopf bifurcation leads to an oscillatory magnetoconvection at the secondary instability. Both the modes $W_{101}$ and $W_{011}$ start periodically varying and show relaxation oscillation.  The solid and broken blue (black) curves marked with small circles show the extrema of the modes $W_{101}$ and $W_{011}$, respectively, for $r_{O1} \le r < 1.077$. The mode $W_{101}$ oscillates between two non-zero positive values. The lower broken blue (black) curve shows that the minimum of the mode $W_{011}$ for the oscillatory solution is always zero in this range of $r$. The blue (black) orbit [Fig.~\ref{fig:vert_phase_DNS}(a)] shows a limit cycle in the $W_{101}-W_{011}$ plane corresponding to an oscillatory solution for $r=1.07$ and $Q=5$. The invariance of the system under a rotation by an angle of $\pi/2$ about a vertical axis gives another oscillatory solution, which may be obtained by the transformation $W_{101} \rightarrow W_{011}$ and  $W_{011} \rightarrow W_{101}$. The solid and broken blue (black) curves in Fig.~\ref{fig:vert_bif_DNS_Q5} would then mean the extrema of the Fourier modes $W_{011}$ and $W_{101}$, respectively, for the second oscillatory solution. The corresponding limit cycle for $r=1.07$ and $Q=5$ is shown by the green (gray) orbit in  Fig.~\ref{fig:vert_phase_DNS}(a). The minimum of one of the modes $W_{101}$ and $W_{011}$ is always zero for $r_{O1} \le r \le r_{H1}$. A set of rolls disappears for a period for which the corresponding roll mode remains at zero. The fluid patterns show the phenomenon of pattern bursting. The mirror symmetries of the flow allow two such limit cycles in each quadrant of the $W_{101}-W_{011}$ plane for $r_{O1} \le r < 1.077$. Depending on the initial conditions, one set of limit cycle is chosen in a simulation. We have shown here the possible limit cycles in the first quadrant of the $W_{101} - W_{011}$ plane. We have labeled the the periodic bursting of patterns as PB-II because two sets of oscillatory solutions exist in each quadrant of the $W_{101}-W_{011}$ plane.

Two limit cycles glue together into one at $r=r_{H1}=1.077$. The temporal variations of both the largest modes $W_{101}$ and $W_{011}$ then become identical with a fixed phase difference between them. We observe identical set of extrema for both the modes $W_{101}$ and $W_{011}$ for $r_{H1} < r < 1.170$. The pink (light gray) curves in Fig.~\ref{fig:vert_bif_DNS_Q5} show the variation of the identical extrema for both the modes with $r$ in the window $r_{H1} < r < 1.170$. Fig.~\ref{fig:vert_phase_DNS}(b) shows a glued limit cycle at $r=1.11$, which touches both the axes in the $W_{101}-W_{011}$ plane for a finite time. The fluid patterns show the phenomena of periodic bursting and are labeled as periodic bursting (PB-I) here.  The minima of both the modes begin to grow continuously and become non-zero [see the lower pink (light gray) curve in Fig.~\ref{fig:vert_bif_DNS_Q5}], as $r$ is raised slowly. There is a continuous transition from a state of glued periodic bursting (PB-I) to a state of glued oscillations oscillating (OCR-I). We have named them differently as the fluid patterns appear different in the two cases. During the phenomenon of the bursting (PB-I), one set of rolls is observed for an interval slightly less than half the period of oscillation and another set of rolls perpendicular to the earlier set for slightly less than another half of the period of oscillation. In the OCR-I state, on the other hand, a pattern of oscillating cross-rolls is always observed  with $|W_{101}|_{max} > |W_{011}|_{max}$ in the first half and $|W_{011}|_{max} > |W_{101}|_{max}$ in the second half of the period of oscillation. A glued limit cycle for an OCR-I state continues to exist for $r_{H1} \le r < 1.170$. The glued limit cycle spontaneously breaks into two limit cycles (OCR-II) at $r=r_{H2}=1.170$. Fig.~\ref{fig:vert_phase_DNS}(c) displays two sets of smaller limit cycles at $r=1.190$. They glue once again at $r=r_{H3}=1.200$. Fig.~\ref{fig:vert_phase_DNS}(d) shows such a state at $r=1.210$. At $r=r_{H4}=1.218$, the glued limit cycle breaks again into two much smaller limit cycles. With further increase in $r$, the smaller limit cycles shrink to two fixed points points corresponding to patterns of stationary cross-rolls (CR) through an inverse Hopf bifurcation at $r=r_{O2}=1.228$. Two sets of cross-rolls become a pattern of stationary squares at $r=r_{PF}=1.241$ via an inverse pitch-fork bifurcation.  
 
However, a little increase in the magnetic field drastically affects the sequence of bifurcations.
The vertical magnetic field delays the onset of convection. The threshold for onset of convection $Ra_c (Q)$ increases with a increase in $Q$. The primary instability appears again in the form of straight rolls. The vertical magnetic field also stabilizes straight rolls and they remain stable upto a larger value of $r$. Fig.~\ref{fig:vert_bif_DNS_Q6} shows a bifurcation diagram  in the presence of a vertical magnetic field for $Q=6$. The straight rolls now exists for $r < 1.31$. The mode $W_{011}$ remains zero for $r < 1.31$. There are again four sets of stationary rolls are possible.  Our choice of random positive initial conditions led to the selection of a pattern of straight rolls parallel to the $y$-axis.  The stationary straight rolls bifurcate directly to a set of stationary squares at the secondary instability, which occurs at $r= 1.31$. The mode $W_{101}$ falls to a much smaller value and the mode $W_{011}$ grows and become equal to the mode $W_{101}$, as shown by black curve in Fig.~\ref{fig:vert_bif_DNS_Q6}.  The stationary squares become unstable at $r=1.36$ via a forward Hopf bifurcation and a time periodic pattern is observed. The temporal variation of the Fourier modes $W_{101}$ and $W_{011}$ are identical ($|W_{101}|_{max}=|W_{011}|_{max}$) but with a constant phase difference between them. The resulting patterns appear as oscillatory cross-rolls (OCR) due to a constant phase difference between two modes. We have labeled therefore these patterns as oscillating cross-rolls (OCR). 

\begin{figure*}[ht]
\begin{center}
\resizebox{0.8\textwidth}{!}{%
  \includegraphics{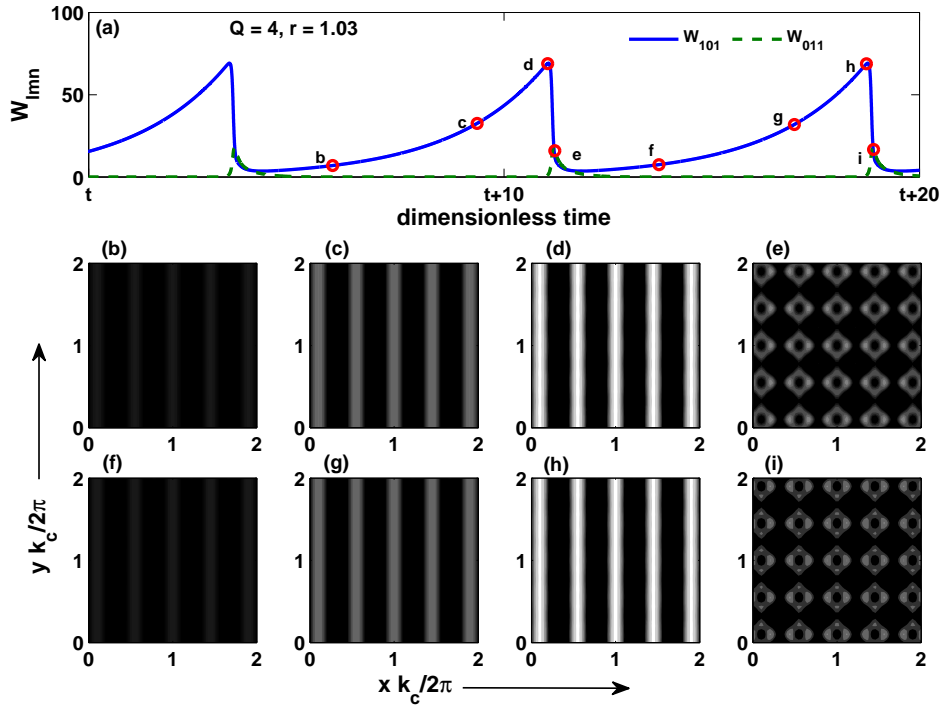}
}
\caption{\label{fig:horiz_PB} (a) Temporal evolution of the two largest Fourier modes $W_{101}$ [blue (black) curve] and $W_{011}$ [green (gray) curve] computed from DNS for a horizontal magnetic field are for $Q = 4$, $r = 1.03$ and $Pr = 0.01$. Several contour plots [(b)-(i)] of the convective temperature field at the mid-plane ($z=1/2$) are shown  for the instants marked by `b' to `i', respectively, in the part (a). The mode $W_{011}$ is excited every time the $W_{101}$ is a little away from its maximum value.}
\end{center}
\end{figure*}

\begin{figure}[h]
\begin{center}
\resizebox{0.45\textwidth}{!}{%
  \includegraphics{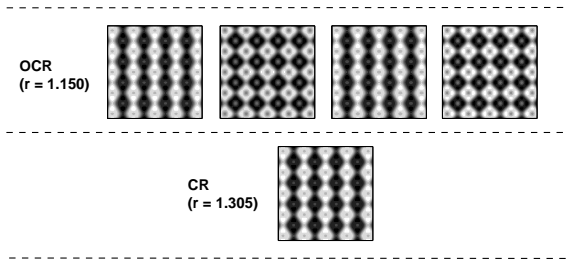}
}
\caption{\label{fig:horiz_patterns} Typical patterns obtained from DNS for horizontal magnetic field for $Q=4$ and $Pr=0.01$.}
\end{center}
\end{figure}

The phenomenon of bursting in a dynamical system was first observed in a low-dimensional model derived for zero-Prandtl-number thermal convection~\cite{kft_1996,kpf_2006}, which proposed a mechanism for the  saturation of thermal convection in the limit of vanishing Prandtl number. Every time the amplitude of a growing straight rolls became large enough, a wave was excited along the roll axis. The  excitation of waves made the rolls wavy (three dimensional) for a short time, which stopped the further growth of rolls. The amplitude crashed to almost zero shortly after the excitation of the waves. The waves disappeared as the roll amplitude rapidly decreased and became infinitesimally small. The straight rolls started growing again till the next burst of waves, which occurred irregularly in time. The bursts changed the flow direction randomly. The phenomenon of bursting was also observed in a low-Prandt-number Rayleigh-B\'{e}nard experiment in  presence of a uniform rotation~\cite{bajaj_etal_2002}. The flow direction was not found to flip in this experiment. The direct numerical simulations (DNS) of zero-Prandtl-number thermal convection with slow rotation (Taylor number $Ta < 100$) showed the possibility of periodic as well as random bursting~\cite{maity_kumar_2014} of flow patterns without change in flow directions. However, the bursts led to reversal of the flow directions at higher rotation rates ($Ta \ge 100$). We have observed only periodic bursting behavior without flow reversal in DNS in the case of magnetoconvection for smaller values of $Q$. In addition, the bursting behavior involves two sets of rolls rather than one set of rolls as observed in the simple models~\cite{kft_1996,kpf_2006}. The bursting behavior is not observed at higher values of $Q$ for the fluid parameters we have used. We are not aware of bursting behavior in an experiment on RB magnetoconvection. However, bursting behavior is reported in fluid dynamos~\cite{gdf_2008}. An experiment on a fluid dynamo by Gallet et al~\cite{gallet_etal_2012} showed a glued limit cycle with flow reversals between two cross-roll states.  

\subsection{\label{sec:patterns_horizontal}Fluid patterns with a horizontal magnetic field}

\begin{figure}[h]
\begin{center}
\resizebox{0.5\textwidth}{!}{%
  \includegraphics{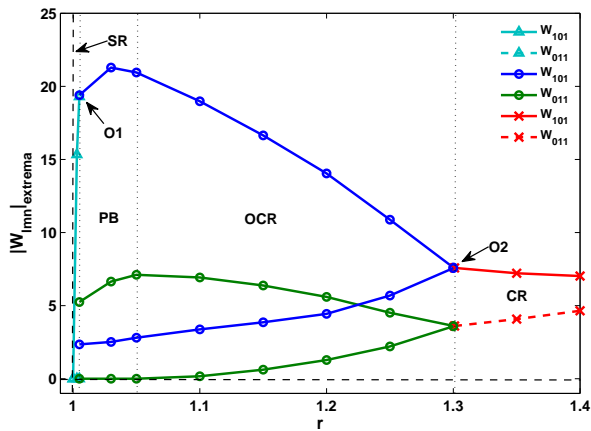}
}
\caption{\label{fig:horiz_bif_DNS_Q10} Bifurcation diagram in presence of a horizontal magnetic field for $Q=10$. The extrema of the two largest Fourier modes $W_{101}$ and $W_{011}$ computed from DNS are plotted as a function of $r$ for $Pr=0.01$. The symbols $O1$ and $O2$ stand for Hopf and inverse Hopf bifurcation points.}
\end{center}
\end{figure}

Now we discuss the effects of a horizontal magnetic field on the convective patterns. The primary instability is unaffected in the presence of a small horizontal magnetic field. Stationary straight rolls observed at the instability onset are always aligned to the horizontal magnetic field. Rolls perpendicular to the direction of the applied magnetic field are not found at the primary instability. The threshold for the secondary instability is very close to the threshold for the primary instability for low-Prandtl-number fluids. For $Pr=0.01$ and $Q=10$, the patterns of straight rolls become oscillatory due to a forward Hopf bifurcation at $r = r_{O1} = 1.007$. We observe that the amplitude of the rolls aligned to the  magnetic field begins increasing exponentially. Shortly before it reaches a critical amplitude, a new set of rolls normal to the direction of the magnetic field is excited. This causes the amplitude of the earlier set of rolls to decrease rapidly to a much smaller value. The amplitude of the new set of rolls starts decreasing and becomes zero, and with this the amplitude of the rolls aligned to the magnetic field  starts growing again. Fig.~\ref{fig:horiz_PB}(a) displays the temporal evolution of the two largest modes $W_{101}$ and $W_{011}$ for $Q=4$ and $r=1.03$. The mode $W_{101}$ shows relaxation oscillation involving two time scales, while the mode $W_{011}$ shows the phenomenon of bursting. As the magnetic field is applied along the $y$-axis, the largest mode $W_{101}$ corresponding to a set of rolls aligned to the magnetic field shows a large variation between two non-zero values.  Contour plots of the temperature field at $z=1/2$ display the patterns [Fig.~\ref{fig:horiz_PB}(b)-(i)] at the instants marked by letters `b' to `i' in Fig.~\ref{fig:horiz_PB}(a). Fluid patterns symmetric under the transformation  $x \rightarrow y$ and $y \rightarrow -x$ are not possible even for a weak horizontal magnetic field. The fluid pattern consists of a set of growing rolls and it appears as squares or cross-rolls for a short period, when both the roll modes have finite values.  The pattern dynamics shows a time periodic bursting near onset. With an increase in $r$ for a fixed value of $Q$, the convection patterns appear as oscillating cross-rolls (OCR) followed by a pattern of stationary cross-rolls (CR). Homoclinic gluing is not observed in this case. Typical flow patterns at secondary and tertiary instabilities are shown in Fig.~\ref{fig:horiz_patterns} for $Q=4$.

Figure~\ref{fig:horiz_bif_DNS_Q10} displays a bifurcation diagram in the presence of a horizontal magnetic field ($Q=10$) applied along the $y$-axis. The variation of the stationary values of the modes $W_{101}$ and $W_{011}$ with respect to $r$ is shown by solid and broken cyan (light gray) curves, respectively. The mode $W_{101}$ grows faster, while the mode $W_{011}$ remains vanishingly small for $r < 1.007$.  The primary instability appears in the form of straight rolls parallel to the direction of the applied magnetic field. As the horizontal magnetic field breaks the rotational symmetry, we never observe a set of rolls perpendicular to the direction of the applied magnetic field.  As $r$ is raised even slightly above $r= r_{O1}=1.007$, we observe both the modes $W_{101}$ and $W_{011}$ begin to oscillate. The blue (black) and green (gray) curves show the variation of the extrema of the roll modes $W_{101}$ and $W_{011}$ with $r$, respectively. The maximum of the mode $W_{101}$ is always larger than the maximum of the mode $W_{011}$. The mode $W_{101}$ oscillates between two non-zero values, while the mode $W_{011}$ oscillates between zero and a non-zero value for $r < 1.1$. The mode $W_{011}$ is excited only for a short interval compared to the periodicity of the mode $W_{101}$. We observe the phenomenon of periodic bursting (PB) very close to the onset. A set of rolls parallel to the magnetic field always exist but another set of rolls perpendicular to the direction of the magnetic field is excited only for a short time. As the invariance of the system under a finite rotation about a vertical axis is broken by a horizontal magnetic field, the flow pattern by with $W_{101} \rightarrow W_{011}$ and $W_{011} \rightarrow W_{101}$ is not possible. However, a similar flow structures with $W_{101} \rightarrow -W_{101}$ and  $W_{011} \rightarrow \pm W_{101}$ are possible. As $r$ is increased further, the minimum of the mode becomes  nonzero and the bursting behavior disappears. The fluid pattern then always appears as oscillating cross-rolls (OCR). The transition from PB state to OCR state is continuous. They are qualitatively the same flow states. The OCR-limit cycles show the phenomenon of bursting when a part of the limit cycle coincide with the $W_{101}$-axis in the $W_{101}-W_{011}$ plane. We have named them differently as the flow structures appear differently. As there is only one limit cycle in any quadrant of the $W_{101}-W_{011}$ plane and there is no saddle fixed point, a homoclinic gluing is not possible in this case. At $r = r_{O2}= 1.3$, the oscillating cross-rolls bifurcate to stationary cross-rolls via an inverse Hopf bifurcation. The solid and broken red (dark gray) curves [Fig.~\ref{fig:horiz_bif_DNS_Q10}] show the variation of the stationary values of the modes $W_{101}$ and $W_{011}$, respectively, with respect to $r$ in the OCR state.

\begin{figure*}[ht]
\begin{center}
\resizebox{0.9\textwidth}{!}{%
  \includegraphics{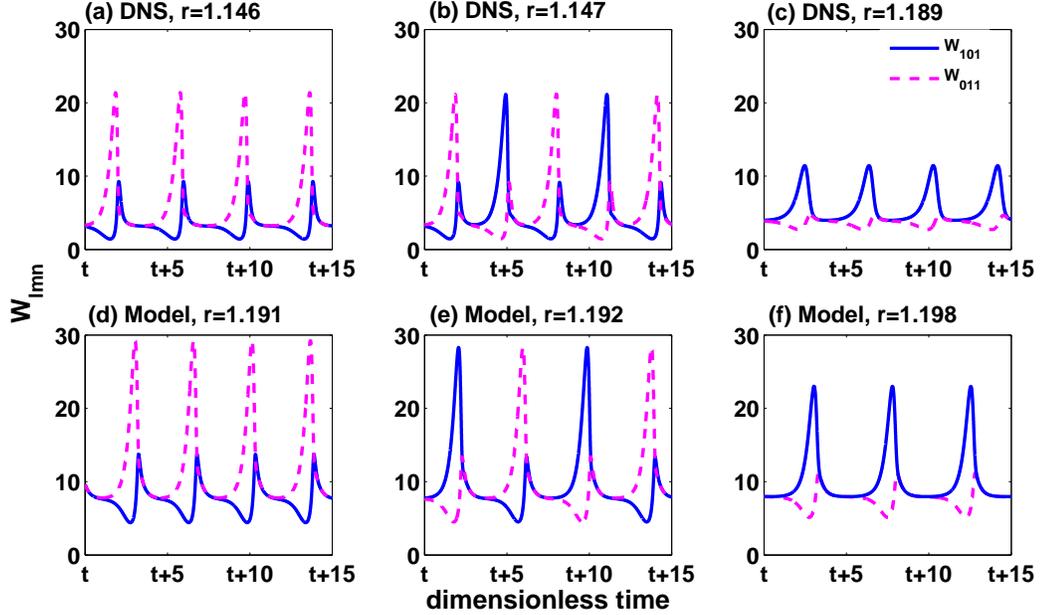}
}
\caption{\label{fig:vert_compare} Comparison of the two dominant Fourier modes $W_{101}$ [blue (black) solid curves] and $W_{011}$ [pink (light gray) dashed curves] obtained from DNS and Model-I for a vertical magnetic field for $Pr=0.01$ and $Q=4$ ($k_c=2.395$). Curves in the upper row [(a), (b) and (c)] are computed from DNS, and the curves in the lower row [(d), (e) and (f)] are computed from Model-I at different values of $r$. We observe qualitatively similar dynamics in both the cases as $r$ is raised in small steps.}
\end{center}
\end{figure*}
 
\section{\label{sec:Model}Low Dimensional Models}
Direct numerical simulations have shown the possibility of the phenomenon of bursting. Locations and stability properties of all fixed points are required to understand the unfolding of bifurcations near the instability onset. The classification of the gluing of two limit cycles into one or breaking of a limit cycle into two requires the knowledge of all the unstable fixed points in the phase space. In addition, DNS for thermal convection at very small values of $Pr$ require a large number of modes even near onset. The details of the unfolding of bifurcations, which is very sensitive to small changes in $Q$, $r$ and $Pr$, require a reasonably good spatial resolution in DNS even very close to the instability onset. This demands enormous amount of computer time for $64^3$ spatial grid points due to critical-slowing-down in the close vicinity of the primary instability.  We therefore  construct low-dimensional models: one for the case of a vertical magnetic field and another for the case of a horizontal magnetic field. The most energetic Fourier modes observed in DNS are used for constructing low-dimensional models. These models although coarse help understanding the unfolding of bifurcations qualitatively near the instability onset.

\subsection{Vertical Magnetic Field}
We expand all the convective fields compatible with the boundary conditions and the symmetries of the system as:
\begin{eqnarray}
v_3(x,y,z,t) &=& [\tilde{W}_{101}\cos{(k_cx)}+\tilde{W}_{011}\cos{(k_cy)} \nonumber \\
&+& \tilde{W}_{121}\cos{(k_cx)}\cos{(2k_cy)} \nonumber \\
&+& \tilde{W}_{211}\cos{(2k_cx)}\cos{(k_cy)}]\sin{(\pi z)} \nonumber \\
&+& \tilde{W}_{112}\cos{(k_cx)}\cos{(k_cy)}\sin{(2\pi z)}
\end{eqnarray}
\begin{eqnarray}
\omega_3(x,y,z,t) &=& \tilde{Z}_{110}\sin{(k_cx)}\sin{(k_cy)} \nonumber \\
&+& \tilde{Z}_{220}\sin{(2k_cx)}\sin{(2k_cy)} \nonumber \\
&+& [\tilde{Z}_{121}\sin{(k_cx)}\sin{(2k_cy)} \nonumber \\
&+& \tilde{Z}_{211}\sin{(2k_cx)}\sin{(k_cy)}]\cos{(\pi z)} \nonumber \\
&+& \tilde{Z}_{112}\sin{(k_cx)}\sin{(k_cy)}\cos{(2\pi z)}
\end{eqnarray}
\begin{eqnarray}
\theta(x,y,z,t) &=& [\tilde{\Theta}_{101}\cos{(k_cx)}+\tilde{\Theta}_{011}\cos{(k_cy)} \nonumber \\
&+& \tilde{\Theta}_{121}\cos{(k_cx)}\cos{(2k_cy)} \nonumber \\
&+& \tilde{\Theta}_{211}\cos{(2k_cx)}\cos{(k_cy)}]\sin{(\pi z)} \nonumber \\
&+& \tilde{\Theta}_{112}\cos{(k_cx)}\cos{(k_cy)}\sin{(2\pi z)} \nonumber \\
&+& \tilde{\Theta}_{002}\sin{(2\pi z)}
\end{eqnarray}

\begin{figure}[ht]
\begin{center}
\resizebox{0.45\textwidth}{!}{%
  \includegraphics{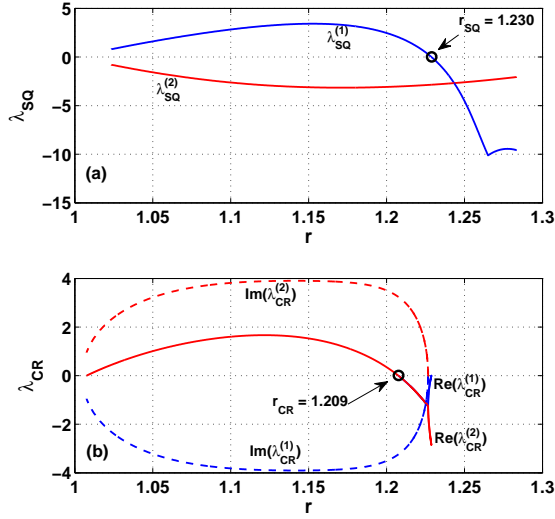}
}
\caption{\label{fig:eigen_SQ_CR_vert} Variations of the eigenvalues of the stability matrix for the fixed points of the model with $r$ for $Pr=0.01$ and $Q=4$ ($k_c=2.395$) for the case of the applied magnetic field in the vertical direction. (a) The blue (black) and red (gray) curves show the variations of the largest eigenvalue $\lambda_{SQ}^{(1)}$ and the second largest eigenvalue $\lambda_{SQ}^{(2)}$ for the square fixed point with  $r$. The stable stationary squares (SQ) are stable for $r > 1.23$. (b) Eigenvalues $\lambda_{CR}^{(1)}$ and $\lambda_{CR}^{(2)}$ of the cross-roll fixed points with the largest real parts form a complex conjugate pair. The solid  and dashed curves show the variations of the real and imaginary parts  of these two eigenvalues with $r$ showing stable stationary cross rolls (CR) for $1.209 < r < 1.23$.}
\end{center}
\end{figure}

\begin{figure}[ht]
\begin{center}
\resizebox{0.5\textwidth}{!}{%
  \includegraphics{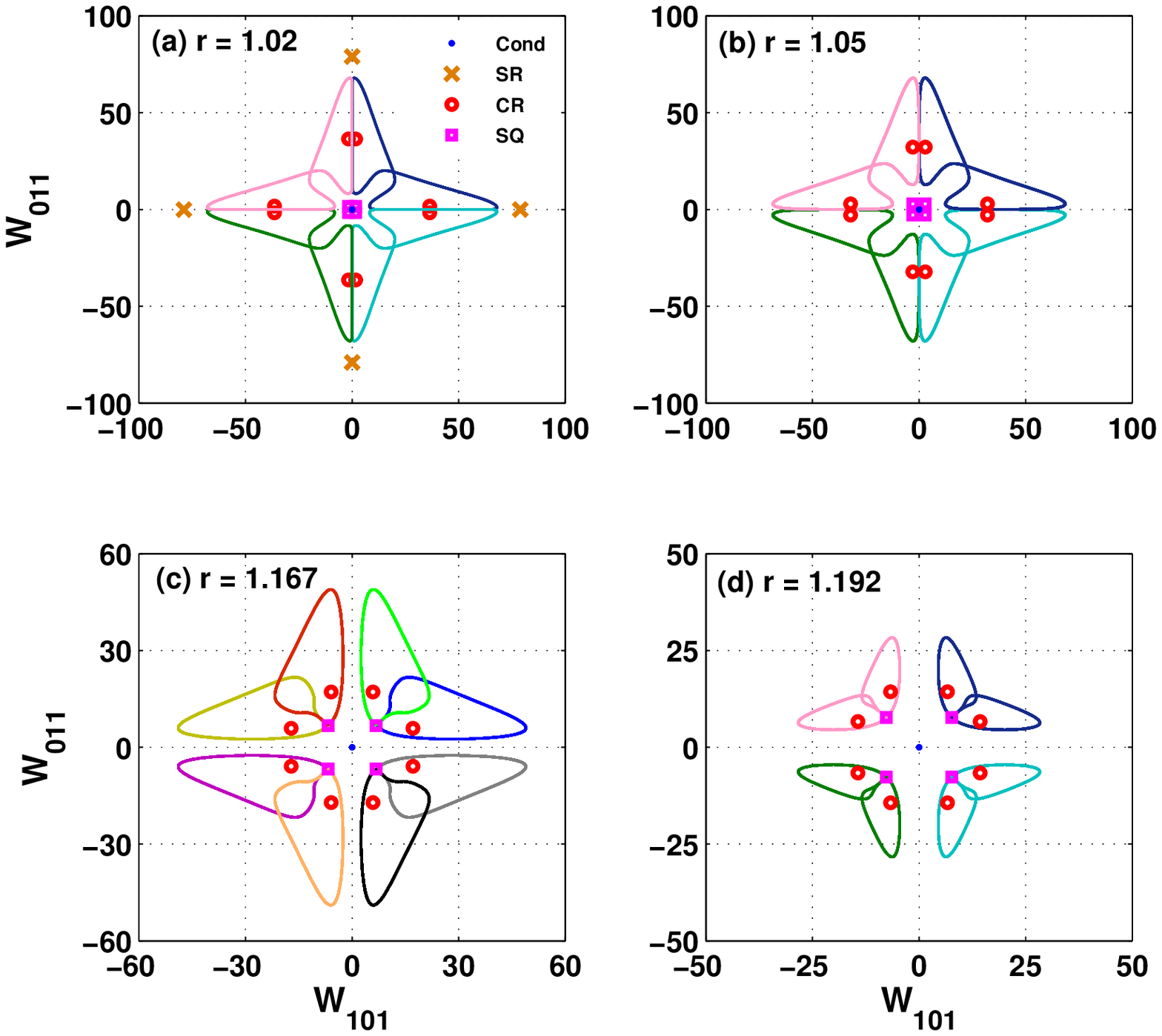}
}
\caption{\label{fig:homoclinic_orbits} Phase portraits in the $W_{101}-W_{011}$ plane as obtained from Model-I showing a glued limit cycle in each quadrant near onset of the secondary instability  for $Q=4$. (a) $r=1.02$, (b) $r=1.05$, (c) $r=1.167$, (d) $r=1.192$. }
\end{center}
\end{figure}

Removing any mode from Model-I shows dynamical behavior qualitatively very different from that observed in DNS. The real modes ($\tilde{W}_{lmn}$) considered in the model are proportional to the corresponding complex modes ($W_{lmn}$) used in DNS. For example, $\tilde{W}_{101} = 2 \times W_{101}$. As the magnetic field $\bm{b}$ is slaved to the velocity field $\bm{v}$ in the limit $Pm \rightarrow 0$, the components of the magnetic field may be expressed in terms of the velocity modes. Projection of the hydrodynamic equations on these Fourier modes leads to a minimum-mode model for a small vertical magnetic field, which is labeled as Model-I. The resulting dynamical system is then integrated by the standard RK4 method using MATLAB, with a time step of $10^{-3}$ or $10^{-4}$. 

\begin{figure}[ht]
\begin{center}
\resizebox{0.5\textwidth}{!}{%
  \includegraphics{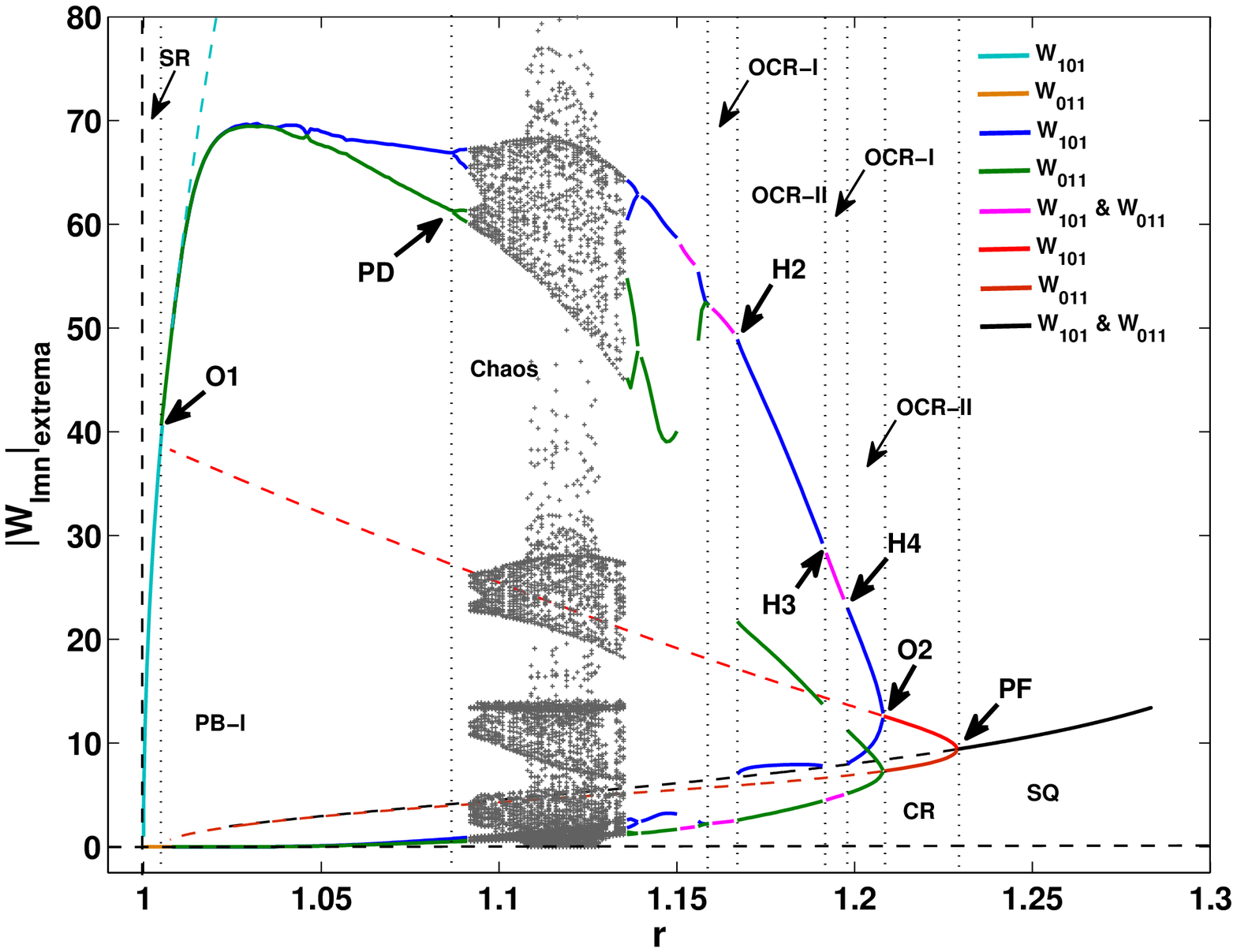}
}
\caption{\label{fig:vert_bif} Bifurcation diagram obtained from Model-I in presence of a vertical magnetic field. The extrema ($|W_{101}|_{extrema}$) of the Fourier mode  $W_{101}$  is plotted as a function of $r$ for $Q=4$ [$k_c (Q) = 2.395$ and $Pr=0.01$].}
\end{center}
\end{figure}

\begin{figure}[ht]
\begin{center}
\resizebox{0.45\textwidth}{!}{%
  \includegraphics{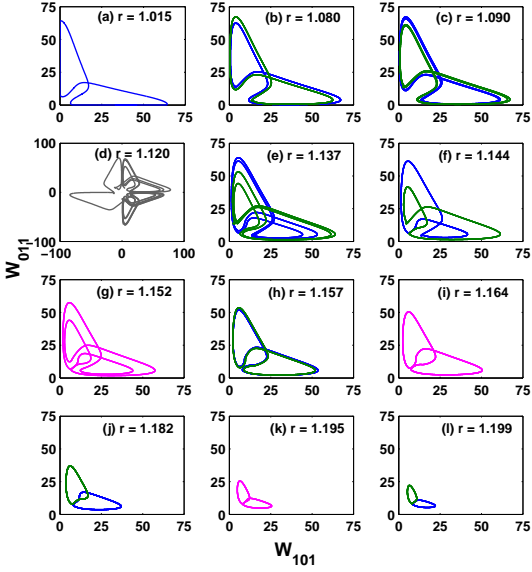}
}
\caption{\label{fig:vert_phase} Phase portraits in the $W_{101}-W_{011}$ plane for different values of $r$ obtained from Model-I for $Q=4$. Other parameters are: $k_c (Q) = 2.395$ and $Pr = 0.01$.}
\end{center}
\end{figure}

\begin{figure}[h]
\begin{center}
\resizebox{0.45\textwidth}{!}{%
  \includegraphics{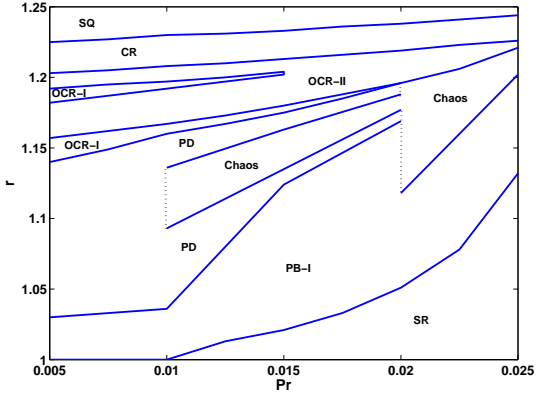}
}
\caption{\label{fig:vert_r_Pr} Fluid patterns in the $r-Pr$ plane for $Q=4$ [$k_c(Q) = 2.395$], as obtained from Model-I. Different regions of the plane show the possibility of straight rolls (SR), periodic bursting (PB-I), bursting with period doubling (PD), chaotic bursting (Chaos), oscillatory cross-rolls with 
$|W_{101}|_{max} = |W_{011}|_{max}$ (OCR-I), oscillatory cross-rolls with 
$|W_{101}|_{max} \neq |W_{011}|_{max}$ (OCR-II), stationary cross rolls (CR) and stationary squares (SQ).}
\end{center}
\end{figure}

\begin{table}[ht]
  \begin{center}
\def~{\hphantom{0}}
  \begin{tabular}{c|c|c|c|c|c|c}
\hline
Bifurcation  &\multicolumn{3}{c|}{$Q=4$}&\multicolumn{3}{c}{$Q=5$} \\
\cline {2-4} \cline{5-7}
points & DNS & Model & Error & DNS & Model & Error \\ 
\hline\hline
$r_{O1}$ & $1.004$ & $1.005$ & $0.10\%$ & $1.011$ & $1.019$ & $0.79\%$ \\
\hline
$r_{H1}$ & $1.056$ & $-$ & $-$ & $1.077$ & $-$ & $-$ \\
\hline
$r_{H2}$ & $-$ & $1.167$ & $-$ & $1.170$ & $1.198$ & $2.39\%$ \\
\hline
$r_{H3}$ & $1.147$ & $1.192$ & $3.92\%$ & $1.200$ & $1.210$ & $0.83\%$ \\
\hline
$r_{H4}$ & $1.189$ & $1.198$ & $0.76\%$ & $1.218$ & $1.211$ & $0.57\%$ \\
\hline
$r_{O2}$ & $1.205$ & $1.209$ & $0.33\%$ & $1.228$ & $1.218$ & $0.81\%$ \\
\hline
$r_{PF}$	& $1.227$ & $1.230$ & $0.24\%$ & $1.241$ & $1.234$ & $0.56\%$ \\
\hline
\hline
 \end{tabular}
\caption{Comparison of bifurcation points computed from DNS and Model-I for a vertical magnetic field for $Pr=0.01$.}
\label{tab:DNS_model_compare}
 \end{center}
 \end{table}

We first compare the two largest Fourier modes $W_{101}$ and $W_{011}$ obtained from the model to those computed from DNS in Fig.~\ref{fig:vert_compare} for $Q = 4$ and $Pr = 0.01$. The Fourier modes $W_{101}$ [blue (black) solid curve] and $W_{011}$ [pink (light gray) dashed curve] represent a set of straight rolls parallel to the $y$-axis and another set of straight rolls parallel to the $x$-axis, respectively. View-graphs in the upper row show the temporal variations of the modes $W_{101}$ and $W_{011}$ computed from DNS and those in the lower row give temporal variations of the same modes obtained from Model-I.  The variations of both the modes obtained from DNS and Model-I appear similar with almost equal time period at $r=r_{H3}$. The values of bifurcation points obtained from Model-I are compared with those computed from DNS in Table~\ref{tab:DNS_model_compare} for $Q=4$ and $Q=5$. The values for various bifurcation points obtained from the model match well with those computed from DNS. The threshold for the secondary bifurcation $r_{O1}$, which is oscillatory, differs less than $1.0\%$.  The model shows the phenomenon of periodic bursting involving a single (glued) limit cycle (PB-I), as expected from DNS for $Q=4$. However, the glued limit cycle obtained from the the model goes under a sequence of period doubling bifurcations instead of spontaneously breaking into two smaller limit cycles. Once the system comes out of the chaotic state at $r=1.153$, it shows  a sequence of bifurcations as observed in DNS for $Q=4$. The maximum error in the various thresholds obtained from the model is less than $4\%$. The amplitudes obtained from the model are  much larger than those computed from DNS near the bifurcation points. It is due to the truncation of modes in the model. The difference in the time periods obtained from the model and DNS increases, as $r$ is raised further. The model, although being coarse, gives additional information about all the relevant fixed points of the system near the instability onset. The first column of Fig.~\ref{fig:vert_compare} shows one of the two possible sets of oscillating cross-rolls (OCR-II) with $|W_{101}|_{max}<|W_{011}|_{max}$. The curves in Fig.~\ref{fig:vert_compare}(a) are computed from DNS at $r = 1.146$ and those in Fig.~\ref{fig:vert_compare}(d) are obtained from Model-I at $r=1.191$. Another set of oscillating cross-rolls may be obtained  by the symmetry $W_{101} \rightarrow W_{011}$ and $W_{011} \rightarrow W_{101}$. The second column of the Fig.~\ref{fig:vert_compare} shows the identical temporal variation for the two modes having a constant phase difference between them.  This case corresponds to a homoclinic gluing of two limit cycles into one, which is symmetric about the diagonal in the $W_{101}-W_{011}$ plane. This is observed  at $r=1.147$ [Fig.~\ref{fig:vert_compare}(b)] in DNS and at $r=1.192$ [Fig.~\ref{fig:vert_compare}(e)] in the model. The glued limit cycle breaks into two smaller limit cycles again at  $r=1.189$ in DNS [Fig.~\ref{fig:vert_compare}(c)] and at $r=1.198$ in Model-I [Fig.~\ref{fig:vert_compare}(f)]. The smaller limit cycles are not symmetric about the diagonal in the $W_{101}-W_{011}$ plane.

Model-I also shows the various stationary patterns, as observed in DNS. They are fixed points of the model. Results of the linear stability analysis of the cross-rolls and squares, computed from the model, are summarized in Fig.~\ref{fig:eigen_SQ_CR_vert}. There are three types of fixed points in Model-I in addition to the trivial fixed point: roll fixed points (SR, $W_{101} \neq 0$, $W_{011}=0$ or $W_{101} = 0$, $W_{011} \neq 0$), square fixed points (SQ, $|W_{101}| = |W_{011}|$) and cross-roll fixed points (CR, $|W_{101}| \neq |W_{011}|$). Two largest eigenvalues of the stability matrix for the square fixed points are plotted as a function of $r$ in Fig.~\ref{fig:eigen_SQ_CR_vert}(a) for $Pr =0.01$ and $Q = 4$. The largest eigenvalue $\lambda_{SQ}^{(1)}$ is always positive and the second largest eigenvalue $\lambda_{SQ}^{(2)}$ is always negative for $0 < r < 1.23$. So the stationary square patterns are actually saddle points for $0 < r < 1.23$. For $r > 1.23$, the two largest eigenvalues of square fixed points become negative.  Figure~\ref{fig:eigen_SQ_CR_vert}(b) shows two largest eigenvalues of the stability matrix for the cross-roll fixed points. They form a complex conjugate pair, with their real parts positive for $1.01 < r < 1.209$. The next largest eigenvalue (not shown here) is negative in this range of $r$. So the cross-roll fixed points are saddle foci for $1.01 < r < 1.209$. The real parts of the complex conjugate pair become negative for $1.209 \le r < 1.23$. So, stationary cross-rolls are stable for $1.209 \le r < 1.23$. For $r \ge 1.23$, the largest real part of the complex conjugate pair becomes positive. Stationary square patterns are therefore stable for $r \ge 1.23$. According to the model, there is no stable stationary pattern for $1.02 < r < 1.209$. The time dependent patterns are observed for $1.01 < r < 1.209$.

Figure~\ref{fig:homoclinic_orbits}(a) shows all the fixed points and trajectories of time dependent solutions computed from the model just above the instability onset ($r=1.02$) for $Q=4$ and $Pr = 0.01$ in the $W_{101}-W_{011}$ plane. The origin, marked by a blue (black) dot, represents the conduction state, which is an unstable node (UN). The origin is surrounded by four saddle points [square fixed points marked by pink (light gray) squares] and eight saddle foci [cross-roll fixed points marked by red (gray) circles]. The saddle points are located along the diagonals ($|W_{101}| = |W_{011}|$)  and the saddle foci are located very close to the two axes in the $W_{101}-W_{011}$ plane. Four unstable nodes [roll fixed points marked by orange (gray) crosses] are located on both the axes symmetrically about the origin. Each quadrant of the $W_{101}-W_{011}$ plane has one saddle square fixed point and two saddle foci.  Any trajectory corresponding to a time dependent solution has to negotiate with these fixed points in the phase space. The location of unstable fixed points, specially saddle foci (unstable cross-rolls), forces the trajectory of a closed orbit to remain alternately on one of the two axes in the $W_{101}-W_{011}$ plane for a finite time.  A set of rolls parallel to the $y$-axis ($x$-axis) disappears for the period the trajectory remains on the $W_{101}$-axis ($W_{011}$-axis).  A closed orbit that encloses two unstable foci in any quadrant of the phase plane represents a glued limit cycle showing the phenomenon of periodic bursting (PB-I). The glued limit cycle in this case is not due to a homoclinic bifurcation as the limit cycle is away from any saddle fixed point.  However, the glued orbit shows a non-local bifurcation involving two unstable nodes (conduction state and a set of straight rolls), a saddle point (squares) and two saddle foci (cross-rolls). Each orbit for glued oscillation is confined to any one quadrant in the $W_{101}-W_{011}$ plane. This does not allow the possibility of flow reversals with time.  The flow reversals in this system would indicate a heteroclinic bifurcation rather than a homoclinic bifurcation. The locations of the saddle point, saddle foci and unstable nodes guide a trajectory to be confined in a quadrant or to wander  beyond a quadrant of the phase plane.  Fig.~\ref{fig:homoclinic_orbits}(b) shows a similar glued limit cycle in each quadrant of the $W_{101}-W_{011}$ plane for $r=1.05$.  The roll fixed points move away from the origin much faster, while the saddle fixed points move away from the  origin along the diagonals slowly in the phase plane with increase in $r$. The size of the limit cycle also decreases slowly with an increase in $r$. The locations of the unstable roll fixed points are not shown in Fig.~\ref{fig:homoclinic_orbits}(b), as they have moved farther away from the origin on the two axes of the phase plane. 

As $r$ is raised to $r=r_{H2}=1.167$, a glued limit cycle touches the saddle fixed point and spontaneously break into two [Fig.~\ref{fig:homoclinic_orbits}(c)]. This allows them to stay slightly away from the saddle fixed point. This is an example of homoclinic breaking and it happens for all the four glued limit cycles. With further increase in $r$ two limit cycles become smaller in size and the saddle points come closed to these smaller orbits. When a limit cycle touches once again the a saddle fixed point they glue together. Fig.~\ref{fig:homoclinic_orbits}(d ) shows homoclinic gluing of two orbits in each quadrant of the $W_{101}-W_{011}$ plane at $r=r_{H3}=1.192$. The model is coarse and not accurate enough when the amplitude of the modes become large enough. For example period doubling is observed in the model but not in DNS. Flow reversals are not observed in a homoclinic gluing for small values of $r$ and $Q$. The experiments by Gallet et al.~\cite{gallet_etal_2012} show flow reversals due to heteroclinic gluing, which may be possible for larger values of $r$. 

Figure~\ref{fig:vert_bif} shows the bifurcation diagram obtained from a low-dimensional model (Model-I) for the vertical magnetic field for $Pr = 0.01$ and $Q = 4$. The stationary values of the Fourier modes $W_{101}$ and $W_{011}$ are plotted as a function of $r$.  The cyan (light gray) curve shows the variation of the stationary value of the mode $W_{101}$ with $r$ near the instability onset. The mode $W_{011}$, shown by a solid orange (gray) curve remains zero near the instability onset. The corresponding pattern is of a stationary rolls parallel to the $y$-axis. Another set of rolls with $W_{101}=0$ and $W_{011} \neq 0$ (not shown here) is also possible for appropriate initial conditions. The magnetoconvection appears as a set of stationary straight rolls (SR) at the primary instability.  The broken cyan (light gray) curve shows the existence of unstable rolls for $r \ge 1.005$. The convection becomes oscillatory at $r = r_{O1} = 1.005$ via a forward Hopf bifurcation. The solid blue (black) and green (gray) curves show the variation of the extrema of the modes $W_{101}$ and $W_{011}$ with respect to $r$, respectively. They are very close to each other  and their minima of both the modes remain zero for $r < 1.037$.  The patterns show the phenomenon of periodic bursting (PB-I). One set of rolls is replaced by a new set of rolls aligned perpendicular to the old set. However, there is no homoclinic bifurcation in the model.  The separation between the maxima of the two modes is visible for $r > 1.037$. 
The minima of both the modes remain very close and begin to grow very slowly. The bursting dynamics goes through a sequence of period doubling at $r=r_{PD}=1.087$ followed by region of chaos for $1.092 \le r \le 1.135$. In this region the the flow show is sensitive to initial conditions and show both homoclinic and heteroclinic chaos. With further increase in $r$ the system comes out of the chaotic flow and shows a sequence of inverse period doubling. Finally at $r = 1.157$, it shows a glued limit cycle. The pink (light gray) curves show the variation of extrema of both the modes with $r$ for a glued limit cycle. Both the modes have identical extrema and a phase difference  between them.  The phase portraits in the $W_{101}-W_{011}$ plane for different values of $r$ are shown in Fig.~\ref{fig:vert_phase}(a)-(f). The phase portraits computed from the model for $1.037 \le r \le 1.157$ are not observed in DNS for $Q=4$ and $Pr=0.01$. However, the model shows qualitatively similar behavior as observed in DNS outside this range of $r$.  The bifurcation diagram obtained from the model clearly shows that a homoclinic bifurcation takes place whenever a limit cycle touches the square saddle fixed point in the phase space. After a series of homoclinic bifurcations (OCR-I $\rightarrow$ OCR-II or OCR-II $\rightarrow$ OCR-I), stationary cross-rolls (CR) are observed. The red (gray) and maroon (deep gray) curves show the variation of the stationary values of modes $W_{101}$ and $W_{011}$, respectively for a set of cross-rolls with $W_{101} > W_{011}$. The symmetries of the system suggests eight such cross-roll fixed points for $1.209 \le r \le 1.229$. The broken red (gray) and  
maroon (deep gray) curves stand for unstable cross-roll fixed points. The unstable cross-rolls (saddle focus) exist for $1 \le r < 1.209$. Stationary cross-rolls bifurcate to stationary squares (SQ), as the amplitudes of the two sets of cross-rolls become equal. The solid black curve stand for stable stationary square solutions ($W_{101} = W_{011}$). The unstable squares (saddle fixed points) are shown by broken black curve. Homoclinic bifurcations occur when a limit cycle touches the corresponding square saddle in the phase space.

\begin{figure}[ht]
\begin{center}
\resizebox{0.5\textwidth}{!}{%
  \includegraphics{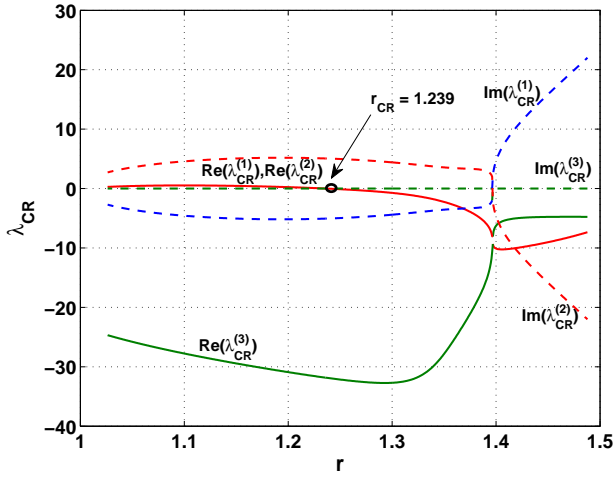}
}
\caption{\label{fig:eigen_CR_horiz} The real and imaginary parts of the three largest eigenvalues ($\lambda_{CR}$) of the stability matrix as a function of $r$ for a horizontal magnetic field ($Q = 4$ and $Pr=0$). Stationary cross-rolls (CR) are found to be stable from the model for $r \ge 1.239$.}
\end{center}
\end{figure}

We now use Model-I to investigate the effects of $Pr$ on fluid patterns for a fixed value of $Q$. Fig.~\ref{fig:vert_r_Pr} shows the possibility of various fluid patterns in different regions of the $r-Pr$ plane for $Q=4$. For $Pr > 0.01$, the patterns at the primary instability are straight rolls. However for very low Prandtl-number fluids ($Pr \le 0.01$) and for a weak magnetic field, the magnetoconvection shows periodic bursting at the primary instability. As $r$ is raised, the phenomenon of bursting goes through a sequence of period doubling to chaotic dynamics. For $Pr > 0.01$, the bursting of patterns appear at the secondary instability. As $r$ is raised in small steps for a small value of $Q=4$ for $Pr > 0.01$, the model shows a series of bifurcations: straight rolls (SR), periodic bursting (PB-I), period doubling (PD), chaotic dynamics, oscillatory behavior (OCR-I, OCR-II), followed again by stationary cross-rolls (CR) and square patterns (SQ). 

\subsection{Horizontal Magnetic Field}
For the case of horizontal magnetic field, we consider the case $Pr \rightarrow 0$ to build a low-dimensional model. In this limiting case, the convective temperature field $\theta$ is also slaved to the vertical velocity $v_3$. Equation~\ref{eq:theta} then reduces to $\nabla^2\theta = -v_3$. The convective fields are expanded as: 
\begin{eqnarray}
v_3(x,y,z,t) &=& [\tilde{W}_{101}\cos{(k_cx)}+\tilde{W}_{011}\cos{(k_cy)} \nonumber \\
&+& \tilde{W}_{121}\cos{(k_cx)}\cos{(2k_cy)} \nonumber \\
&+& \tilde{W}_{211}\cos{(2k_cx)}\cos{(k_cy)}]\sin{(\pi z)} \nonumber \\
&+& \tilde{W}_{112}\cos{(k_cx)}\cos{(k_cy)}\sin{(2\pi z)}
\end{eqnarray}
\begin{eqnarray}
\omega_3(x,y,z,t) &=& \tilde{Z}_{110}\sin{(k_cx)}\sin{(k_cy)} \nonumber \\
&+& \tilde{Z}_{112}\sin{(k_cx)}\sin{(k_cy)}\cos{(2\pi z)}
\end{eqnarray}

We have chosen a few largest modes observed in DNS. The modes $\tilde{Z}_{220}$, $\tilde{Z}_{121}$ and $\tilde{Z}_{211}$ used in Model-I are found to be much smaller compared to the other modes. They are therefore dropped here.  The hydrodynamic equations are now projected on these modes to obtain a seven-mode dynamical system. Dropping any mode further at this stage changes the dynamical behavior observed in the model from that observed in  DNS.  However, the model after adiabatic elimination~\cite{pal_etal_2013} of the linearly decaying Fourier modes $\tilde{W}_{112}$, $\tilde{Z}_{112}$ and $\tilde{Z}_{110}$ preserves the qualitative behavior observed in DNS. The resulting model then involves a set of four coupled nonlinear ordinary differential equations with $r$ and $Q$ as control parameters. This minimum-mode model in the presence of a horizontal magnetic field is called as Model-II in this article. 

In the limit of vanishing Prandtl number, the growing straight rolls are the exact solution of the nonlinear system~\cite{kft_1996}. There is no stationary straight rolls in this case. Stationary square patterns are also not possible, as the horizontal magnetic field breaks the rotational symmetry of the problem. The only non-trivial fixed points are stationary cross-rolls. Results of the linear stability analysis around the fixed points corresponding to cross-rolls are summarized in Fig.~\ref{fig:eigen_CR_horiz}. The eigenvalues with the largest real part form a complex conjugate pair with the real part positive for $1 < r < 1.239$ and negative for $r \ge 1.239$. The third largest eigenvalue is always negative. The cross-rolls are saddle foci for $r < 1.239$. Stationary cross-rolls become stable for $r > 1.239$. 

\begin{figure}[ht]
\begin{center}
\resizebox{0.5\textwidth}{!}{%
  \includegraphics{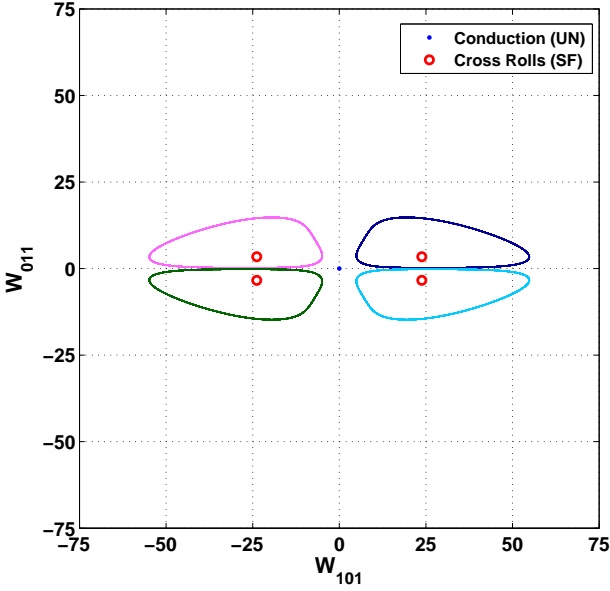}
}
\caption{\label{fig:horiz_phase_orbits} Phase portrait in the $W_{101}-W_{011}$ plane as obtained from Model-II slightly above the onset of primary instability ($r=1.1$) for $Q=4$ and $Pr \rightarrow 0$.}
\end{center}
\end{figure}

Figure~\ref{fig:horiz_phase_orbits} shows the phase portrait, as computed from Model-II, for a weak horizontal magnetic field ($Q=4$) in the limit $Pr \rightarrow 0$ for $r=1.1$. One unstable cross-roll fixed point (saddle focus) is located close to the $W_{101}$ axis in each quadrant of the $W_{101}-W_{011}$ plane. They are connected by the inversion symmetry. Each possible limit cycle encloses only one cross-roll fixed point near the instability onset. The locations of the cross-roll fixed points make a part of all possible limit cycles in the phase plane to coincide with the $W_{101}$-axis for a finite time. The resulting patterns of oscillating cross-rolls therefore display periodic bursting at the primary instability in very low-$Pr$ fluids. As the trajectory of any limit cycle during the phenomenon of bursting interacts with a node at the origin (conduction state) and two saddle foci (unstable cross-roll fixed points) located closely on two sides of the $W_{101}$-axis, a local bifurcation analysis around the conduction state is unlikely to capture this behavior. As there is only one limit cycle in each quadrant and no saddle fixed point in this case, a homoclinic gluing is not possible. 

\begin{figure}[h]
\begin{center}
\resizebox{0.5\textwidth}{!}{%
  \includegraphics{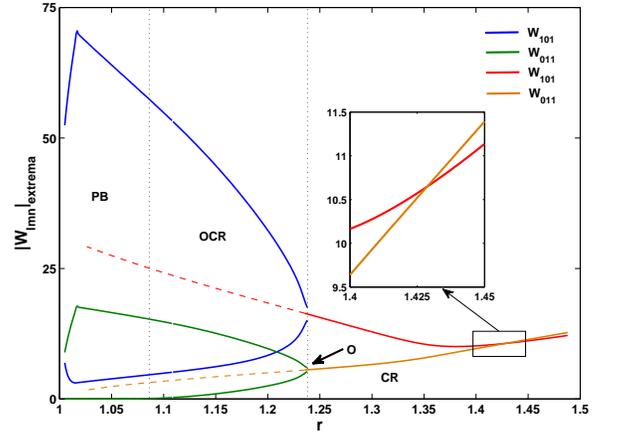}
}
\caption{\label{fig:horiz_bif} Bifurcation diagram obtained from Model-II for a  horizontal magnetic field for $Q=4$ and $Pr=0$. The extrema of the roll modes $W_{101}$ [(blue) curves] and $W_{011}$ [(green) curves] for a time periodic flow are plotted as a function of $r$. Stationary cross-rolls (asymmetric squares) are possible for $r > 1.238$. Inset shows that $W_{101} > W_{011}$ for $r < 1.428$ and $W_{101} < W_{011}$ for $r > 1.428$.} 
\end{center}
\end{figure}

Figure~\ref{fig:horiz_bif} shows the bifurcation diagram in the case of horizontal magnetic field. Blue (black) and green (gray) curves show the variations of the extrema of Fourier modes $W_{101}$ and $W_{011}$ respectively with $r$ for $Q = 4$ in the limit $Pr \rightarrow 0$. The minima of the mode $W_{011}$ remains as zero near the primary instability. The convection shows periodic bursting (PB) instead of stationary straight rolls at the primary instability for $Pr = 0$ to construct Model-II.  The minima of the mode $W_{011}$ begins to grow slowly, as $r$ is increased further. There is a smooth transition from bursting behavior to oscillatory behavior without bursting (OCR). The phenomenon of bursting is also an OCR solution in this case. As the rotational symmetry is broken by the horizontal magnetic field, there is only one set of OCR solution. The limit cycle shrinks to a point in the $W_{101}-W_{011}$  plane at $r =r_{O}= 1.23$ via an inverse Hopf bifurcation. Oscillatory cross-rolls bifurcate to stationary cross-rolls (CR). There is only one set of stationary cross-rolls for $r \ge 1.23$. The solid red (gray) and orange (light gray) curves show the variations of the fixed values of the modes $W_{101}$ and $W_{011}$ with $r$, respectively. The broken red and and orange curves give the variations of the values of the modes $W_{101}$ and $W_{011}$ with $r$ respectively for unstable cross-rolls fixed points.  The inset gives the stationary values of the Fourier modes $W_{101}$ and $W_{011}$ as a function of $r$ for $1.4 < r < 1.45$. The amplitude of a set of rolls aligned to the applied magnetic field is always larger than the same for a set of rolls in the perpendicular direction. A comparison of the threshold for the inverse Hopf bifurcation point obtained from Model-II and DNS shows that errors in the value of $r_O$ obtained from the model is within $5\%$ for $Q=5$ and within $3\%$ for $Q=10$.

\begin{figure}[ht]
\begin{center}
\resizebox{0.45\textwidth}{!}{%
  \includegraphics{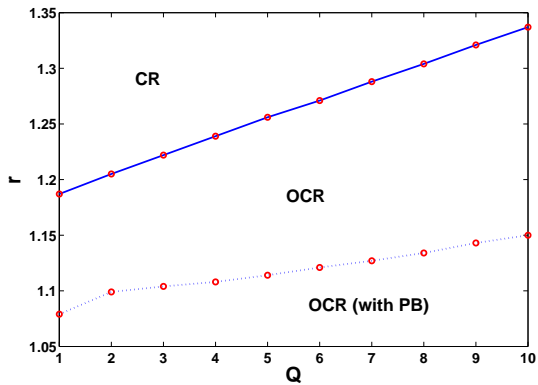}
}
\caption{\label{fig:horiz_r_Q} Fluid patterns in the $r-Q$ plane for $Pr=0$ obtained from Model-II for a weak horizontal magnetic field.  Oscillating cross-rolls (OCR) are observed with and without periodic bursting (PB) in addition to the stationary cross rolls (CR). The maximum of the largest mode $W_{101}$ for rolls parallel to the magnetic field is much larger than that of the largest mode $W_{011}$ for the  set of rolls perpendicular to the magnetic field.}
\end{center}
\end{figure}

Figure~\ref{fig:horiz_r_Q} shows the possible fluid patterns in different regions of the $r-Q$ plane, as obtained from Model-II.  The onset of magnetoconvection appears as oscillating cross-rolls showing periodic bursting (PB) for a weak magnetic field in the limit $Pr \rightarrow 0$. The rolls with larger amplitude are always aligned to the applied magnetic field above the onset. As the amplitude of the growing rolls reaches close to its maximum value, a new set of rolls of much smaller amplitude is excited in the direction normal to the applied magnetic field for a very short time. This brings down the amplitude of the  growing rolls to a smaller non-zero value. As soon as the amplitude of rolls parallel to the magnetic field reaches close to its minimum value, the rolls in the  perpendicular direction disappear. This is different than the dynamics observed in zero-Prandtl-number convection~\cite{pal_etal_2013,thual_1992} in the absence of an external magnetic field ($Q = 0$), where the phenomena of bursting is always chaotic showing flow reversals. As $r$ is raised slightly, the minimum value of the amplitude of new set of rolls becomes non-zero. As a result, the phenomenon of bursting disappears. However, $|W_{101}|_{max}$ is always much larger than $|W_{011}|_{max}$.  We do not observe any homoclinic bifurcation in this case, as the symmetry $W_{101}$ $\leftrightharpoons$ $W_{011}$ is broken by the horizontal magnetic field. When $r$ is raised further, the only possible limit cycle shrinks to a point in the phase space leading to a stationary fluid pattern of cross-rolls (CR). This behavior qualitatively matches with the results obtained from DNS for very small values of $Pr$. 

\section{Conclusions}
We have investigated the effects of small magnetic field on the  homoclinic bifurcations in thermal convection in a low-Prandtl-number fluids ($Pr = 0.01$) using DNS. A series of homoclinic bifurcations occurs near the primary instability in the presence of a weak magnetic field in the vertical direction, as $r$ is raised in small steps. The phenomenon of bursting leads to a periodic replacement of a growing set of rolls after a finite time by a new set of rolls in the direction perpendicular to the older set. Two limit cycles may spontaneously glue into a larger one or a  limit cycle may spontaneously break into two smaller ones at homoclinic bifurcations. As $r$ is raised further, the limit cycle solutions bifurcate to stationary cross-rolls through an inverse Hopf bifurcation and finally to stationary square patterns via an inverse pitchfork bifurcation.

For a small horizontal magnetic field, a set of rolls aligned normal to the magnetic field is excited for a short time at regular intervals of time. This happens in the close vicinity of the primary instability for a low-Prandtl-number fluid. This leads to a sudden fall of the amplitude of rolls aligned along the magnetic field. The horizontal magnetic field breaks the symmetry of the hydrodynamic system under the rotation by $\pi/2$ about a vertical axis and allows only one set of limit cycle. 
A weak horizontal magnetic field acts as a symmetry-breaking perturbation on global bifurcation and it rules out a homoclinic gluing. As $r$ is raised to higher values, the limit cycle solution bifurcates to a set of stationary cross-rolls by inverse Hopf bifurcation. We have also presented minimum-mode models for a weak applied magnetic field, which qualitatively capture the behavior observed in DNS for smaller values of $r$.\\

\section*{Acknowledgment}
We are thankful to both the anonymous Referees for their useful suggestions which have  improved the manuscript.


\end{document}